\documentclass[apj,iop,numberedappendix]{emulateapj}
\usepackage{graphicx}
\usepackage{natbib}
\usepackage{lineno}
\usepackage{amsmath}
\usepackage{amsfonts} 
\usepackage{amssymb}
\usepackage{xspace}

\usepackage[pdftex]{hyperref}
\usepackage[x11names]{xcolor}

\hypersetup{pdftitle={Evidence against star-forming galaxies as the dominant source of IceCube neutrinos},
pdfsubject={Evidence against star-forming galaxies as the dominant source of IceCube neutrinos},
pdfauthor={Keith Bechtol, Markus Ahlers, Mattia Di Mauro, Marco Ajello, and Justin Vandenbroucke},
pdfstartview={FitH},
colorlinks=true,
bookmarksopen=false,
bookmarksnumbered=false,
bookmarksopenlevel=0,
linkcolor=Blue1!60!black,
citecolor=Green1!50!black,
urlcolor=Blue1!70!black
}

\graphicspath{{./}{./figs/}}

\mathchardef\mhyphen="2D

\newlength{\dhatheight}

\newcommand{\sfg}{{SFG}\xspace}
\newcommand{\sfgs}{{SFGs}\xspace}

\newcommand{\gammaRays}{{$\gamma$ rays}\xspace}
\newcommand{\gammaRayHyph}{{$\gamma$-ray}\xspace}

\providecommand\physrep{\ref@jnl{Phys.~Rep.}}%
\providecommand\apjs{\ref@jnl{ApJS}}%
\providecommand{\jcap}{\ref@jnl{JCAP}}%
\begin{document} 
\title{Evidence against star-forming galaxies as the dominant source of\\ IceCube neutrinos}
\shorttitle{Evidence against star-forming galaxies as the dominant source of IceCube neutrinos}

\author{Keith Bechtol\altaffilmark{1,2,\textdagger}}
\author{Markus Ahlers\altaffilmark{1,2,\textdaggerdbl}}
\author{Mattia Di Mauro\altaffilmark{3,4,5}}
\author{Marco Ajello\altaffilmark{6}}
\author{Justin Vandenbroucke\altaffilmark{1,2}}
\altaffiltext{1}{Wisconsin IceCube Particle Astrophysics Center (WIPAC), Madison, WI 53703, USA}
\altaffiltext{2}{Department of Physics, University of Wisconsin--Madison, Madison, WI 53706, USA}
\altaffiltext{3}{W.~W.~Hansen Experimental Physics Laboratory, Stanford University, Stanford, CA 94305, USA}
\altaffiltext{4}{Kavli Institute for Particle Astrophysics and Cosmology, Stanford University, Stanford, CA 94305, USA}
\altaffiltext{5}{Department of Physics and SLAC National Accelerator Laboratory, Stanford University, Stanford, CA 94305, USA}
\altaffiltext{6}{Department of Physics and Astronomy, Clemson University, Clemson, SC 29634, USA}
\altaffiltext{\textdagger}{\href{mailto:keith.bechtol@icecube.wisc.edu}{keith.bechtol@icecube.wisc.edu}}
\altaffiltext{\textdaggerdbl}{\href{mailto:markus.ahlers@icecube.wisc.edu}{markus.ahlers@icecube.wisc.edu}}

\begin{abstract}
The cumulative emission resulting from hadronic cosmic-ray interactions in star-forming galaxies (\sfgs) has been proposed as the dominant contribution to the astrophysical neutrino flux at TeV to PeV energies reported by IceCube. 
The same particle interactions also inevitably create \gammaRayHyph emission that could be detectable as a component of the extragalactic \gammaRayHyph background (EGB), which is now measured with the \textit{Fermi}-LAT in the energy range from 0.1 to 820 GeV. 
New studies of the blazar flux distribution at \gammaRayHyph energies above 50~GeV place an upper bound on the residual non-blazar component of the EGB. 
We show that these results are in strong tension with models that consider \sfgs as the dominant source of the diffuse neutrino backgrounds.
A characteristic spectral index for parent cosmic rays in starburst galaxies of $\Gamma_{\rm SB} \simeq2.3$ for $dN/dE \propto E^{-\Gamma_{\rm SB}}$ is consistent with the observed scaling relation between \gammaRayHyph and IR luminosity for \sfgs, the bounds from the non-blazar EGB, and the observed \gammaRayHyph spectra of individual starbursts, but underpredicts the IceCube data by approximately an order of magnitude.
\end{abstract}

\submitted{}


\maketitle 

\section{Introduction}
\label{sec:intro}

Extragalactic \gammaRays and high-energy neutrinos represent a census of particle acceleration and other non-thermal processes throughout the observable universe. Neutrinos in particular trace the interactions of relativistic nuclei, which are the energetically dominant component of cosmic rays (CRs). The IceCube Collaboration has now measured an astrophysical flux of neutrinos at energies from 10~TeV to $\gtrsim1$~PeV. 
This signal has been detected in various analyses and found to be consistent with an isotropic and equal-flavor flux of neutrinos that is expected from extragalactic source populations. 
The combined best-fit power-law flux in all flavors in the 25~TeV to 2.8~PeV energy range is $E^2\phi(E) = 6.7^{+1.1}_{-1.2} \times 10^{-8}(E/100~{\rm TeV})^{-0.5\pm0.09}~{\rm GeV}~{\rm cm}^{-2}~{\rm s}^{-1}~{\rm sr}^{-1}$ \citep{Aartsen:2015ita}.

There are many proposed candidate sources of TeV to PeV astrophysical neutrinos. 
Extragalactic source candidates include galaxies with intense star formation~\citep{Loeb:2006tw,Stecker:2006vs,Murase:2013rfa,He:2013cqa,Anchordoqui:2014yva,Chang:2014hua,Chang:2014sua,Senno:2015tra,Emig:2015dma}, cores of active galactic nuclei (AGN)~\citep{Stecker:1991vm,Stecker:2013fxa,Kalashev:2014vya}, low-luminosity AGN~\citep{Bai:2014kba,Kimura:2014jba}, blazars~\citep{Tavecchio:2014eia,Padovani:2014bha,Dermer:2014vaa,Padovani:2015mba}, low-power \gammaRayHyph bursts (GRBs)~\citep{Waxman:1997ti,Murase:2013ffa,Ando:2005xi,Tamborra:2015qza}, cannonball GRBs~\citep{Dado:2014mea}, intergalactic shocks~\citep{Kashiyama:2014rza}, and active galaxies embedded in structured regions~\citep{Berezinsky:1996wx,Murase:2008yt,Murase:2013rfa}. 
However, no individual high-energy neutrino sources have been identified yet in a variety of different searches. 
Constraints from up-going track event searches in IceCube~\citep{Aartsen:2014cva,Aartsen:2014ivk} imply that the source population responsible for the observed astrophysical neutrino flux has a density of $\gtrsim10^{-6}$~Mpc$^{-3}$ if the constituents are continuous emitters \citep{Ahlers:2014ioa}.

Given that the same particle interactions that produce high-energy neutrinos also inevitably generate high-energy \gammaRays, multi-messenger studies can provide further insight on the origins of the IceCube signal. For example, the inelastic collisions of CR nucleons with ambient matter in interstellar and intergalactic space create pions whose decay products include energetic \gammaRays and neutrinos.
This process is expected to be the dominant high-energy emission mechanism in star-forming galaxies (\sfgs) and may be relevant for other hadronuclear sources, such as galaxy clusters.
Several authors have jointly considered the cumulative neutrino and \gammaRayHyph emissions of extragalactic source populations in light of recent results from IceCube and the Large Area Telescope (LAT) on board the \textit{Fermi Gamma-ray Space Telescope} (\textit{Fermi}) \citep[e.g.,][]{Murase:2013rfa,Chang:2014hua,Chang:2014sua,Tamborra:2014xia,Ando:2015bva}.
 
In this study, we critically examine the hypothesis that CR induced emission in \sfgs can account for a majority of the astrophysical neutrino flux measured with IceCube. We find that such a scenario is difficult to reconcile with new studies of the extragalactic \gammaRayHyph background (EGB) composition at energies above 50~GeV~\citep{TheFermi-LAT:2015ykq}, which are briefly reviewed in the next section. In Section~\ref{sec:multi-messenger}, we compute the cumulative \gammaRayHyph and neutrino emission expected from the evolving population of \sfgs and compare these fluxes to the \gammaRayHyph and neutrino data. We then consider generic CR calorimeter models in Section~\ref{sec:calorimeter}. General considerations and systematic uncertainties are discussed Section~\ref{sec:systematic}. Finally, we consider the implications of these multi-messenger constraints for the origin of the IceCube signal in Section~\ref{sec:discussion}.

\section{Non-blazar component of the EGB}
\label{sec:nonblazar}

Significant advances have been made in our understanding of the EGB in recent years. The spectrum of the EGB has now been measured with the \textit{Fermi}-LAT in the energy range from 0.1 to 820 GeV \citep{Ackermann:2014usa}. Meanwhile, more than one thousand extragalactic \gammaRayHyph sources have been individually detected, mostly blazars \citep{3FGL,Ackermann:2015yfk}, and multiple source classes are known to contribute to the EGB at varying levels across this broad energy range \citep{Ajello:2015mfa,DiMauro:2015ika}.

Especially rapid progress has been made in the energy range above 10~GeV, where the LAT has unprecedented sensitivity due to a combination of large collecting area ($\sim1$~m$^2$), excellent angular resolution ($\sim0.1$~deg), and high background rejection efficiency. The Second \textit{Fermi} Hard Source List (2FHL) includes 360 sources that are significantly detected at energies above 50~GeV in 80 months of sky-survey data \citep{Ackermann:2015uya}. At high Galactic latitudes ($|b|>10^{\circ}$), the 2FHL catalog is dominated by AGN, which account for $90\%$ of the sources; $70\%$ are associated to specific BL Lac type blazars, and the total blazar fraction is estimated to be $97\%$.

In addition to the individually resolved 2FHL sources, which comprise $\sim40$ percent of the total EGB intensity, the flux distribution of sources fainter than the detection threshold of about $8 \times 10^{-12}$ ph cm$^{-2}$ s$^{-1}$ has been constrained by the statistical distribution of individual photons \citep{TheFermi-LAT:2015ykq}. Specifically, the number of spatial pixels containing varying numbers of photons can provide information of the number of sources at fluxes down to about $1.3 \times 10^{-12}$ ph cm$^{-2}$ s$^{-1}$. The 2FHL catalog sources and pixel counting method together yield a best-fit flux distribution which is well parameterized by a broken power law with a flux break in the range $[0.8,1.5]\times 10^{-11}$ ph cm$^{-2}$ s$^{-1}$ and a slope above and below the break equal to $\alpha_1=2.49$ and $\alpha_2\in[1.60,1.75]$, with $dN/dS \propto S^{-\alpha}$.

\begin{deluxetable}{lcc}
\tablewidth{\columnwidth} 
\tablecaption{Spectral indices of \gammaRayHyph-detected starburst galaxies\label{tab1}}
\tablehead{
\colhead{Name}&
\colhead{Spectral Index}&
\colhead{Energy Range [GeV]}}
\startdata
M82\,${}^{a,b}$&$2.21\pm0.06$&$0.1-100$\\
 &$2.5\pm0.6_{\rm stat}\pm0.2_{\rm stat}$&$700-5\times10^{3}$\\
NGC 253\,${}^c$&$2.34\pm0.03$&$0.2-3\times10^{4}$\\
NGC 4945\,${}^b$&$2.43\pm0.07$&$0.1-100$\\
NGC 1068\,${}^b$&$2.32\pm0.10$&$0.1-100$\\
NGC 2146${}^b$&$2.37\pm0.15$&$0.1-100$\\
Arp 220\,${}^d$&$2.35\pm0.16$&$0.2-100$
\enddata
\tablenotetext{}{${}^a$\,\citep{3FGL} ; ${}^b$\,\citep{Ackermann:2012vca} ; ${}^c$\,\citep{Abramowski:2012xy}  ;  ${}^d$\,\citep{Peng:2016nsx}}
\end{deluxetable}

The integral of this flux distribution is $2.07^{+0.40}_{-0.34}\times10^{-9}$ ph cm$^{-2}$ s$^{-1}$ sr$^{-1}$ compared to the total EGB intensity above 50~GeV of $(2.40\pm0.3)\times10^{-9}$ ph cm$^{-2}$ s$^{-1}$ sr$^{-1}$. In other words, blazars comprise $86^{+16}_{-14}\%$ of the total EGB intensity~\citep{TheFermi-LAT:2015ykq}.\footnote{Point  sources  with  fluxes $S > 1.3 \times 10^{-12}$ ph cm$^{-2}$ s$^{-1}$ produce $1.47^{+0.20}_{-0.24} \times 10^{-9}$ ph cm$^{-2}$ s$^{-1}$ sr$^{-1}$ (61\% of the EGB), while $6.0^{+2.0}_{-1.0} \times 10^{-10}$ ph cm$^{-2}$ s$^{-1}$ sr$^{-1}$ (25\% of the EGB) is produced by sources below that flux ~\citep{TheFermi-LAT:2015ykq}.} The best-fit cumulative intensity of residual emission, from both discrete extragalactic sources and truly diffuse processes, is $14\%$, corresponding to an intensity of $3.3\times10^{-10}$ ph cm$^{-2}$ s$^{-1}$ sr$^{-1}$ above 50~GeV. Taking uncertainties into account, the upper bound for the non-blazar fraction of the EGB is $28\%$ ($6.6\times10^{-10}$ ph cm$^{-2}$ s$^{-1}$ sr$^{-1}$).

\citet{Lisanti:2016jub} performed a similar Non-Poissonian Template Fit (NPTF) of LAT data in the $>50$~GeV energy range and found that point sources account for at least $68^{+9}_{-8}\%$ ($\pm10\%$ systematic uncertainty) of the total EGB intensity.
The NPTF method loses sensitivity to sources below the single-photon limit, corresponding to a flux of $\sim4 \times 10^{-12}$ ph cm$^{-2}$ s$^{-1}$ in the \citet{Lisanti:2016jub} analysis, and therefore represents a lower bound on the point-source contribution, as a realistic source population would include contributions from members of the same population significantly below that flux threshold.
We note that the contribution of sub-threshold sources inferred from the photon fluctuation analysis of \citet{TheFermi-LAT:2015ykq} is consistent with expectations based on blazar luminosity functions \citep{Ajello:2015mfa,DiMauro:2013zfa,Giommi:2015ela}.
Given the consistency between the results of \citet{TheFermi-LAT:2015ykq} and \citet{Lisanti:2016jub} above the single-photon flux threshold, we conclude that an upper bound on the non-blazar EGB fraction of 28\% is reasonable.

Another photon-fluctuation analysis of LAT data in the 1 to 10 GeV energy range has been used by \citet{Zechlin:2015wdz} to constrain the abundance of sources about an order of magnitude fainter than the flux threshold of the 3FGL catalog \citep{3FGL}.
That analysis found that the high-latitude \gammaRayHyph sky ($|b| > 30^{\circ}$) is composed of $(69 \pm 2)\%$ Galactic foreground, $(25 \pm 2)\%$ point sources brighter than $5 \times 10^{-12}$ ph cm$^{-2}$ s$^{-1}$, and $(6 \pm 2)\%$ isotropic diffuse emission (including misclassified CR backgrounds).
These results further support the claim that a majority of the EGB can be attributed point sources.

\section{Cumulative gamma-ray and neutrino flux from SFGs}
\label{sec:multi-messenger} 

The hadronic emission of \sfgs is thought to originate from CR interactions in interstellar space, analogous to the diffuse emission observed from our own Galaxy. The residency time of CRs in a given galaxy is determined by the timescale of diffusive escape, transport by advective outflows, and hadronic interactions with ambient gas. If the loss time is dominated by diffusive escape, the hadronic emission follows a $dN/dE \sim E^{-\Gamma-\delta}$ spectrum where $\Gamma$ is the effective index of the injected CR nucleon spectrum and $\delta$ is the index of the energy dependence of the diffusion tensor. For diffusive shock acceleration, we expect that on average $\Gamma \simeq 2$, although individual accelerators in special environments might have harder spectra \citep{Bykov:2015nta}. Typical values of $\delta$ considered for Galactic CR diffusion are $\delta\simeq1/2$ (Kraichnan) or $\delta\simeq 1/3$ (Kolmogorov). Note that if CRs are accelerated in multiple source populations with different rigidity cutoffs and mass compositions, the resulting effective nucleon spectrum can have additional spectral features.

On the other hand, starburst galaxies, a subset of \sfgs that undergo an episode of vigorous star formation in their central regions, have gas densities that are much higher than observed in quiescent galaxies~\citep{Tacconi:2005nx,Sargent:2012rj}. Diffusion in starburst galaxies might also become weaker due to strong magnetic turbulence~\citep{Thompson:2009jt,Batejat:2011fv}, while advective processes might be enhanced \citep{Lehnert:1996}. Since losses by inelastic collisions and advection are nearly independent of energy, the hadronic emission of starbursts is expected to follow more closely the injected CR nucleon spectrum, $E^{-\Gamma}$. 
Indeed, the nearby starburst galaxies detected at GeV and TeV energies \citep{Acciari:2009wq,Ackermann:2012vca,Abramowski:2012xy,Tang:2014dia,Peng:2016nsx,Griffin:2016wzb} exhibit harder \gammaRayHyph spectral indices than that of the Milky Way and other quiescent galaxies, as summarized in Table~\ref{tab1}. Due to the harder emission and higher pion production efficiency, the starburst subset is predicted to dominate the total diffuse \gammaRayHyph emission of \sfgs beyond a few GeV~\citep{Tamborra:2014xia}. Provided that the CR accelerators in starburst galaxies are capable of reaching per-nucleon energies exceeding $20-30$~PeV, the hadronic emission can also contribute significantly to the diffuse neutrino emission at PeV energies~\citep{Loeb:2006tw}. 

\begin{figure}[t]\centering
\includegraphics[width=0.95\linewidth]{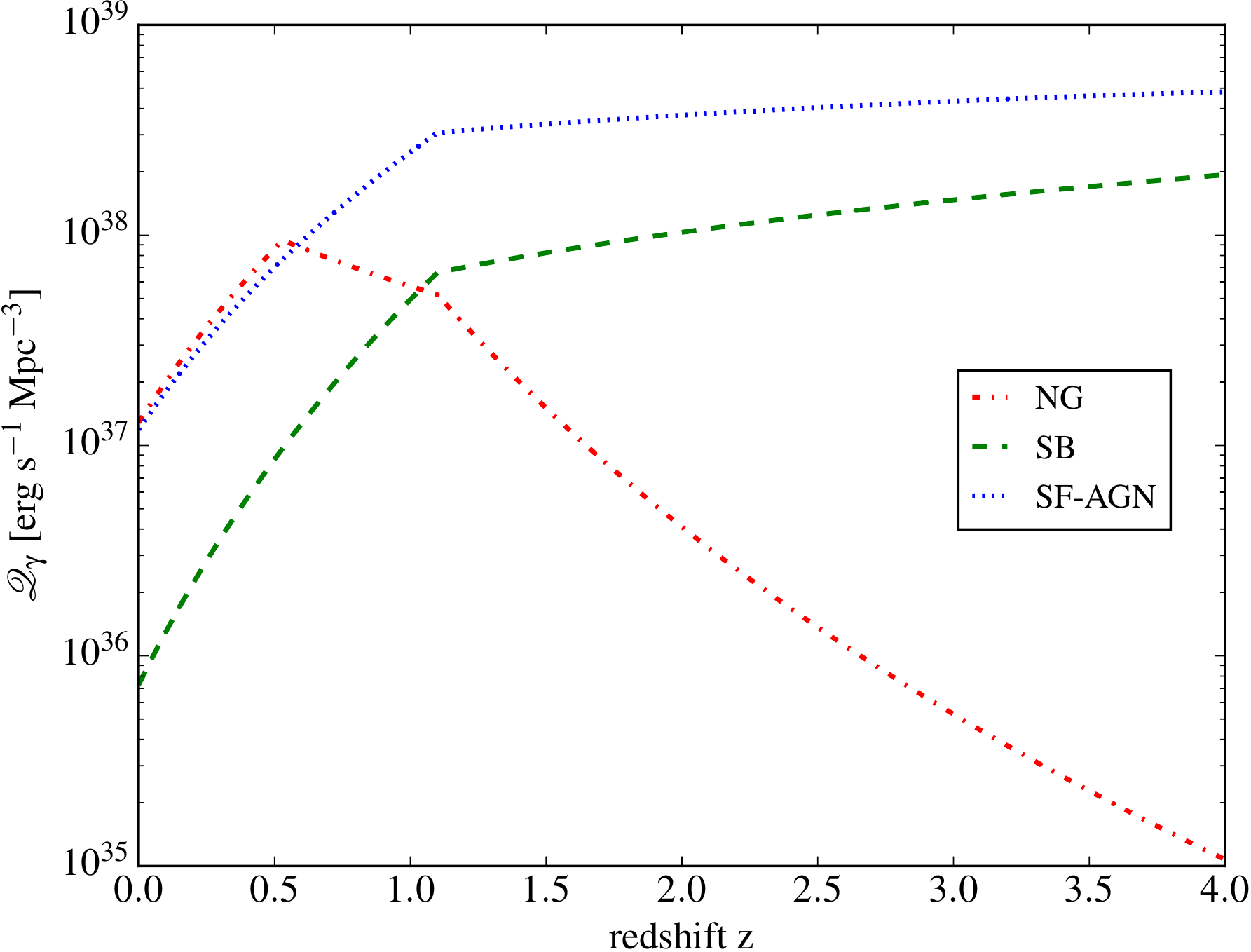}
\caption[]{The \gammaRayHyph luminosity densities of normal galaxies (NG), starburst galaxies (SB), and star-forming galaxies containing an AGN (SF-AGN) following the model of \cite{Tamborra:2014xia}.}\label{fig1}
\end{figure}

\begin{figure*}[t]\centering
\includegraphics[width=0.45\linewidth]{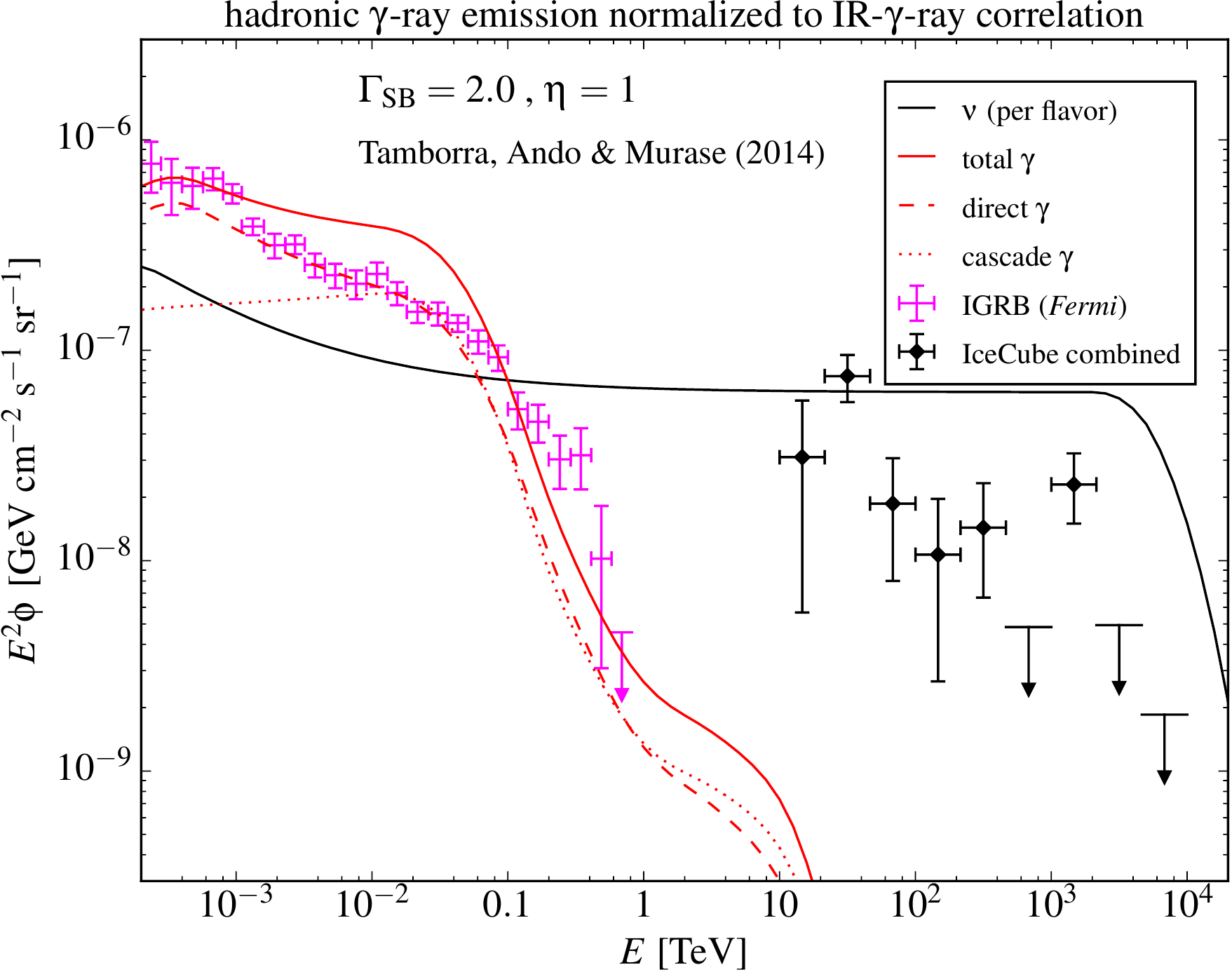}\hspace{0.3cm}
\includegraphics[width=0.45\linewidth]{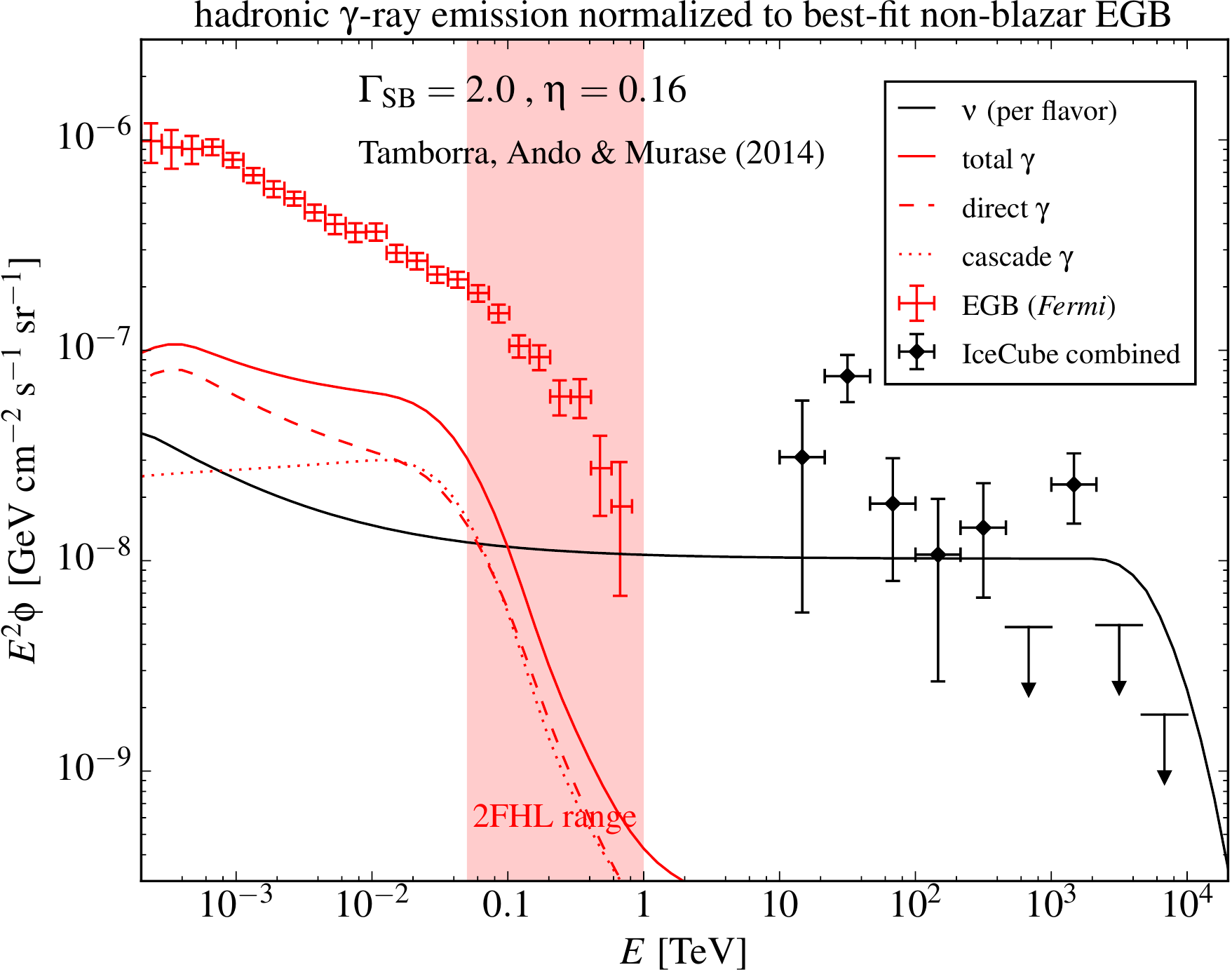}\\[0.3cm]
\includegraphics[width=0.45\linewidth]{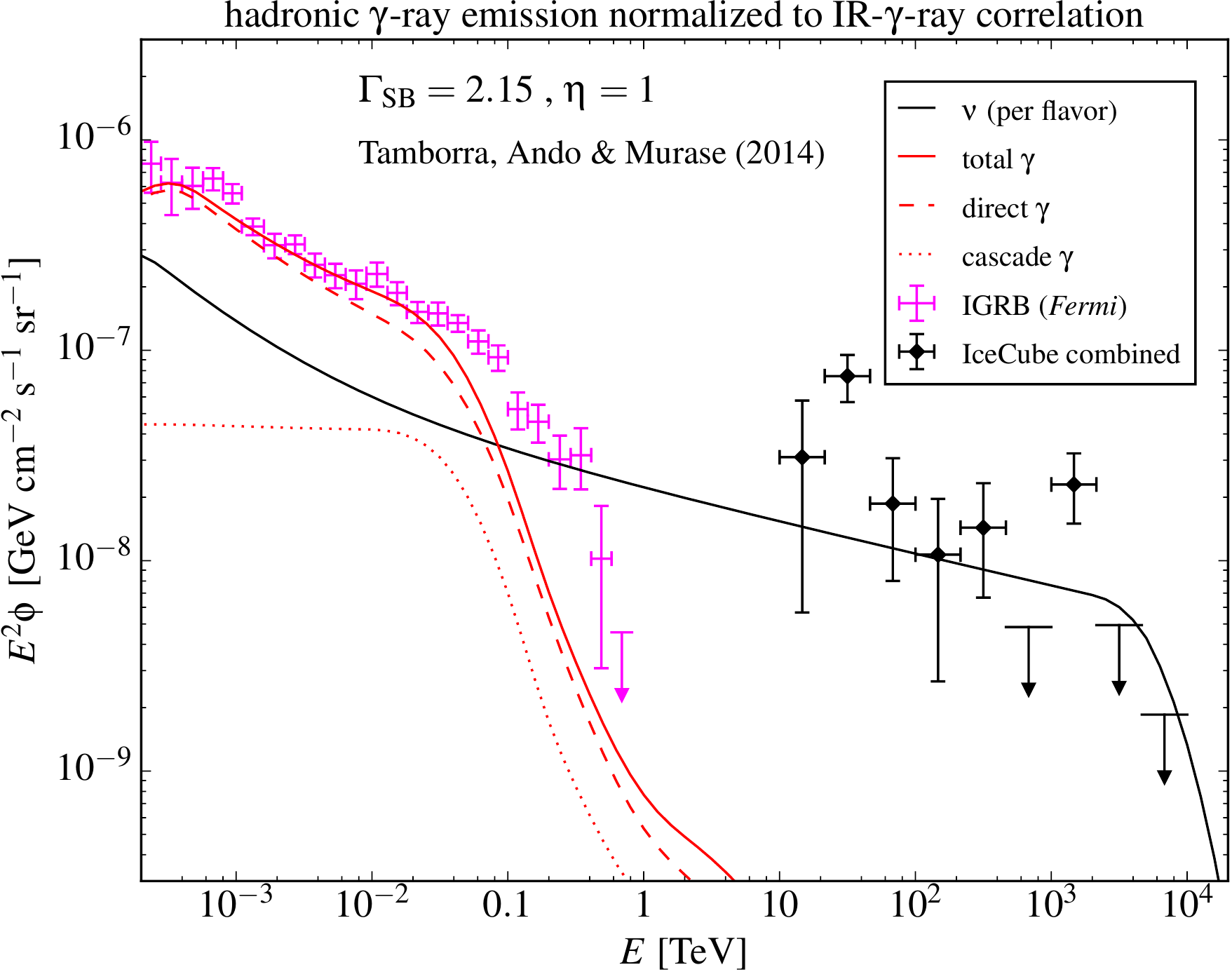}\hspace{0.3cm}
\includegraphics[width=0.45\linewidth]{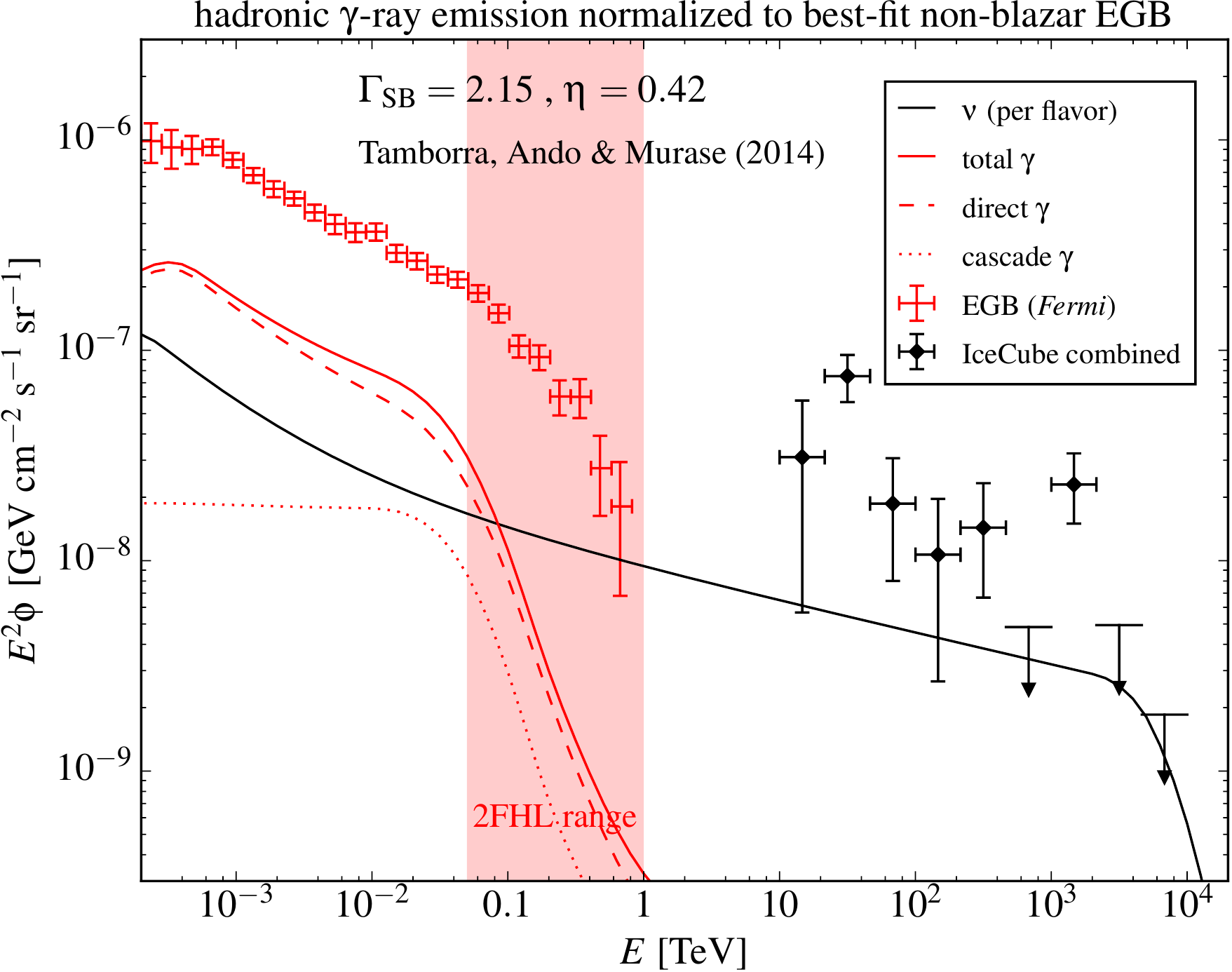}\\[0.3cm]
\includegraphics[width=0.45\linewidth]{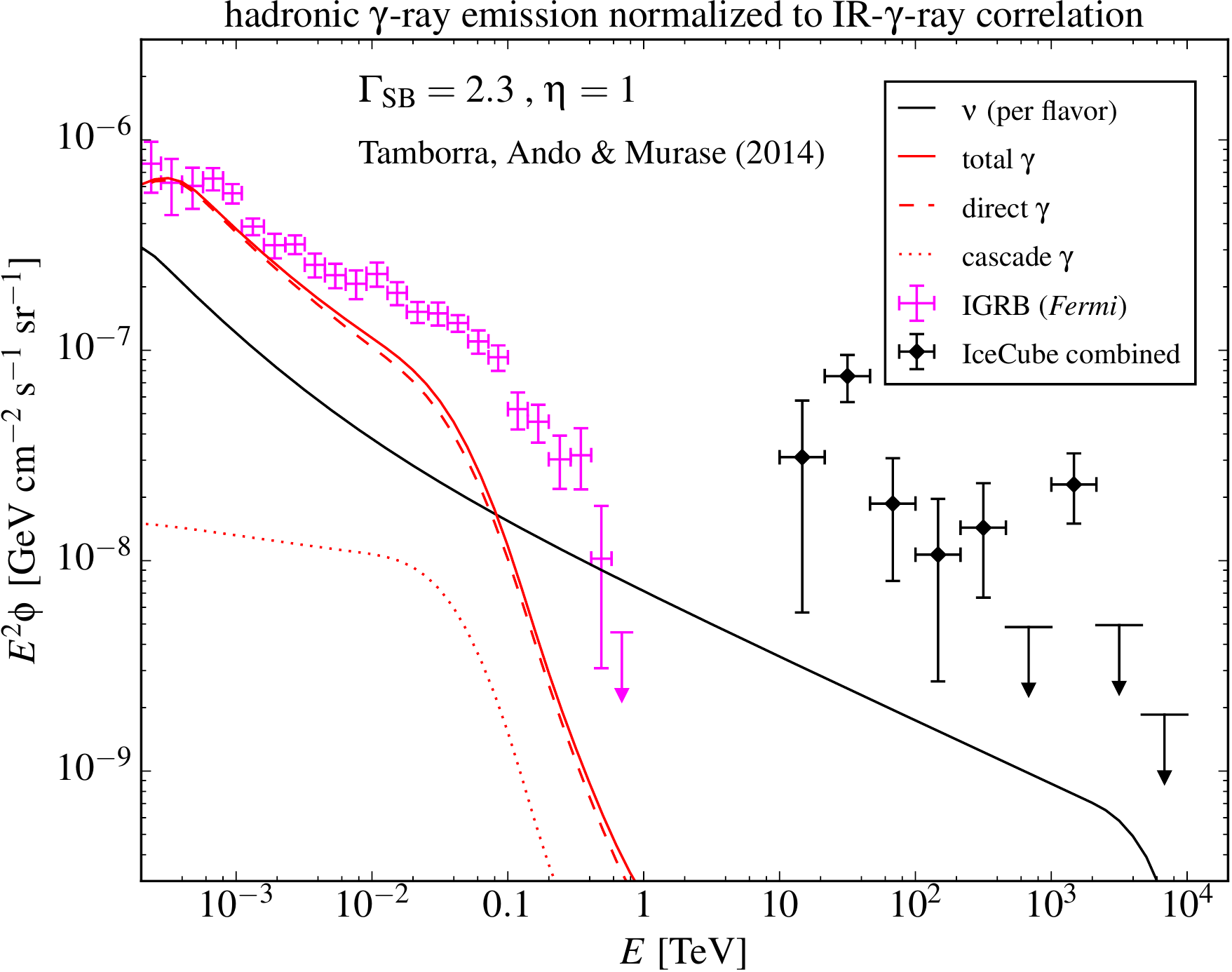}\hspace{0.3cm}
\includegraphics[width=0.45\linewidth]{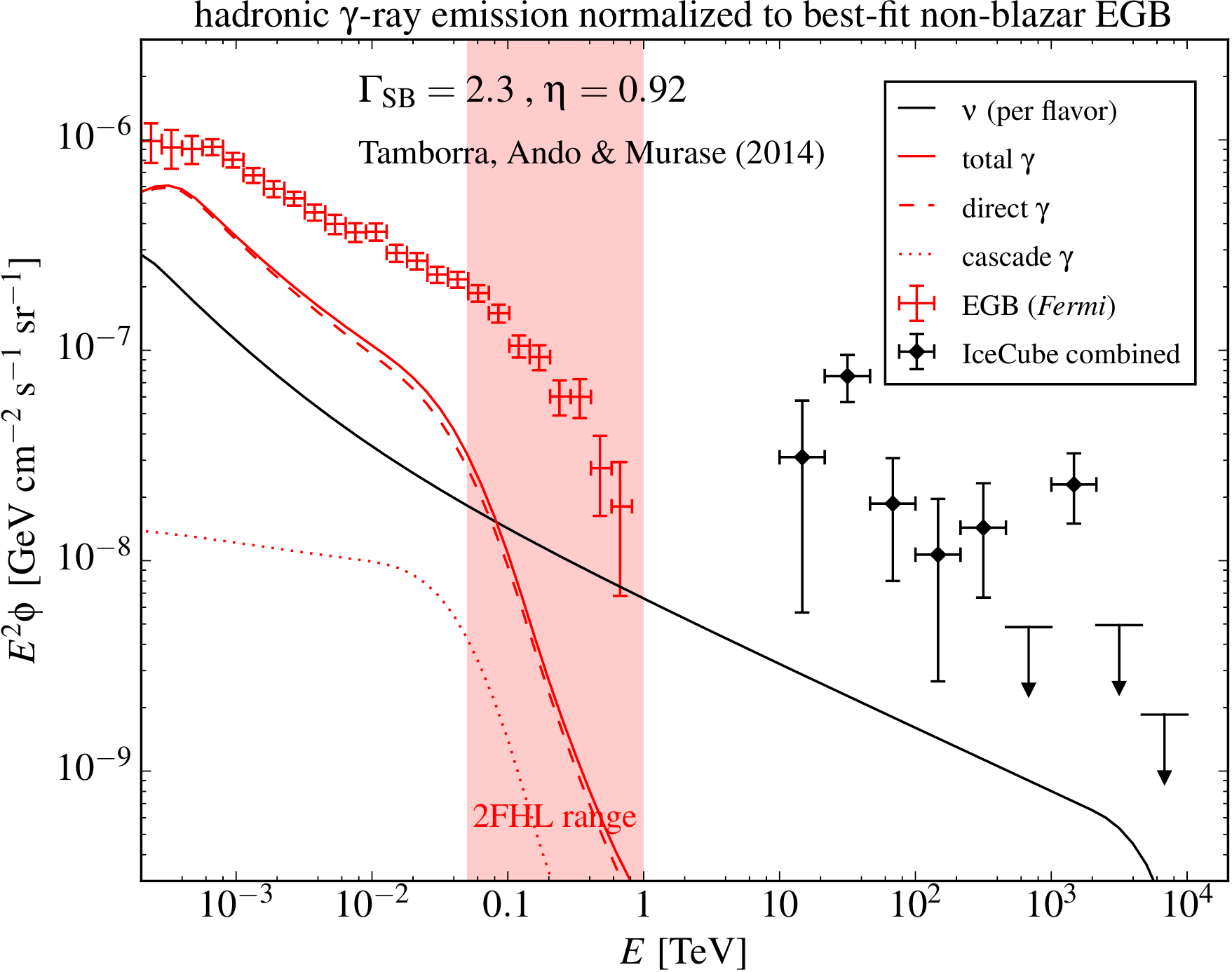}
\caption[]{The \gammaRayHyph (red lines) and per-flavor neutrino (black lines) hadronic emission of \sfgs following the model of \cite{Tamborra:2014xia}. We show the contributions of direct and cascade \gammaRays separately as dashed and dotted lines, respectively. Each row corresponds to a different value for the starburst galaxy spectral index $\Gamma_{\rm SB}$.  In the left panels, the emission is normalized according to the IR-\gammaRayHyph correlation of \sfgs with $\eta = 1$. In the right panels, we show the same model normalized to the best-fit non-blazar EGB emission in the 0.05--1~TeV energy range (red-shaded area).}\label{fig2}
\end{figure*}

In this section we consider hadronic \gammaRayHyph and neutrino production in \sfgs following the model of \cite{Tamborra:2014xia}. In this model the contributions of normal galaxies (NG), starburst galaxies (SB) and star-forming galaxies containing an active galactic nucleus (SF-AGN) are treated with separate luminosity functions and emission spectra. The individual \gammaRayHyph luminosity functions are normalized to the observed infrared (IR) luminosity function from \textit{Herschel} \citep{Gruppioni:2013jna} using the IR-\gammaRayHyph luminosity correlation derived by \cite{Ackermann:2012vca}. 
The \gammaRayHyph emission spectrum for an individual source of population $X$ is assumed to follow 
\begin{equation}\label{eq:NgammaSFG}
\frac{dN_{\gamma,X}}{dE_\gamma} \propto \begin{cases}E_\gamma^{-1.5}&E_\gamma<0.6{\rm GeV}\\E_\gamma^{-\Gamma_X}&0.6{\rm GeV}<E_\gamma<20{\rm PeV}\\E_\gamma^{-\Gamma_X}e^{-E/{\rm 20 PeV}}&20{\rm PeV}<E_\gamma.\end{cases}
\end{equation}
As discussed above, starburst galaxies are expected to have a hard spectral index, $\Gamma_{\rm SB} = \Gamma$, whereas normal galaxies are expected to produce softer emission, $\Gamma_{\rm NG} = \Gamma + \delta$. In our calculations we fix $\delta=1/2$ assuming a Kolmogorov-like energy dependence of CR diffusion. For the case of SF-AGN galaxies we follow the procedure of \cite{Tamborra:2014xia} and divide the population into two sub-populations of NG-like galaxies with index $\Gamma_{\rm NG}$ and SB-like galaxies with index $\Gamma_{\rm SB}$ according to the weighting factors shown in their Table~2. 

After integrating over the IR luminosity distributions of the three populations $X$ one arrives at the \gammaRayHyph luminosity densities $\mathcal{L}_{\gamma,X}$ that are shown in Figure~\ref{fig1}. The \gammaRayHyph emission rate density can then be expressed as
\begin{equation}
Q_\gamma(z,E_\gamma) = \eta\sum_{\rm X}\mathcal{L}_{\gamma,X}(z)\frac{1}{\mathcal{N}_X}\frac{d
{N}_{\gamma,X}(E_\gamma)}{dE_\gamma}\,,
\end{equation}
with normalization\footnote{Our normalization condition differs from that of \cite{Tamborra:2014xia} in that we fix the $\gamma$-ray luminosity in the 0.1--100GeV interval in the source reference frame. However, this has only a negligible effect for the calcuation.}
\begin{equation}
\mathcal{N}_X = \int_{0.1{\rm GeV}}^{100{\rm GeV}} d E_\gamma E_\gamma\frac{d{N}_{\gamma,X}(E_\gamma)}{dE_\gamma}\,.
\end{equation}
In the following, we also allow for a scaling factor $\eta$ in the overall normalization. The model of \cite{Tamborra:2014xia} corresponds to $\eta=1$, based on the observed IR-\gammaRayHyph correlation at $z=0$ and the observed IR luminosity function.

\begin{figure}[t]\centering
\includegraphics[width=0.9\linewidth]{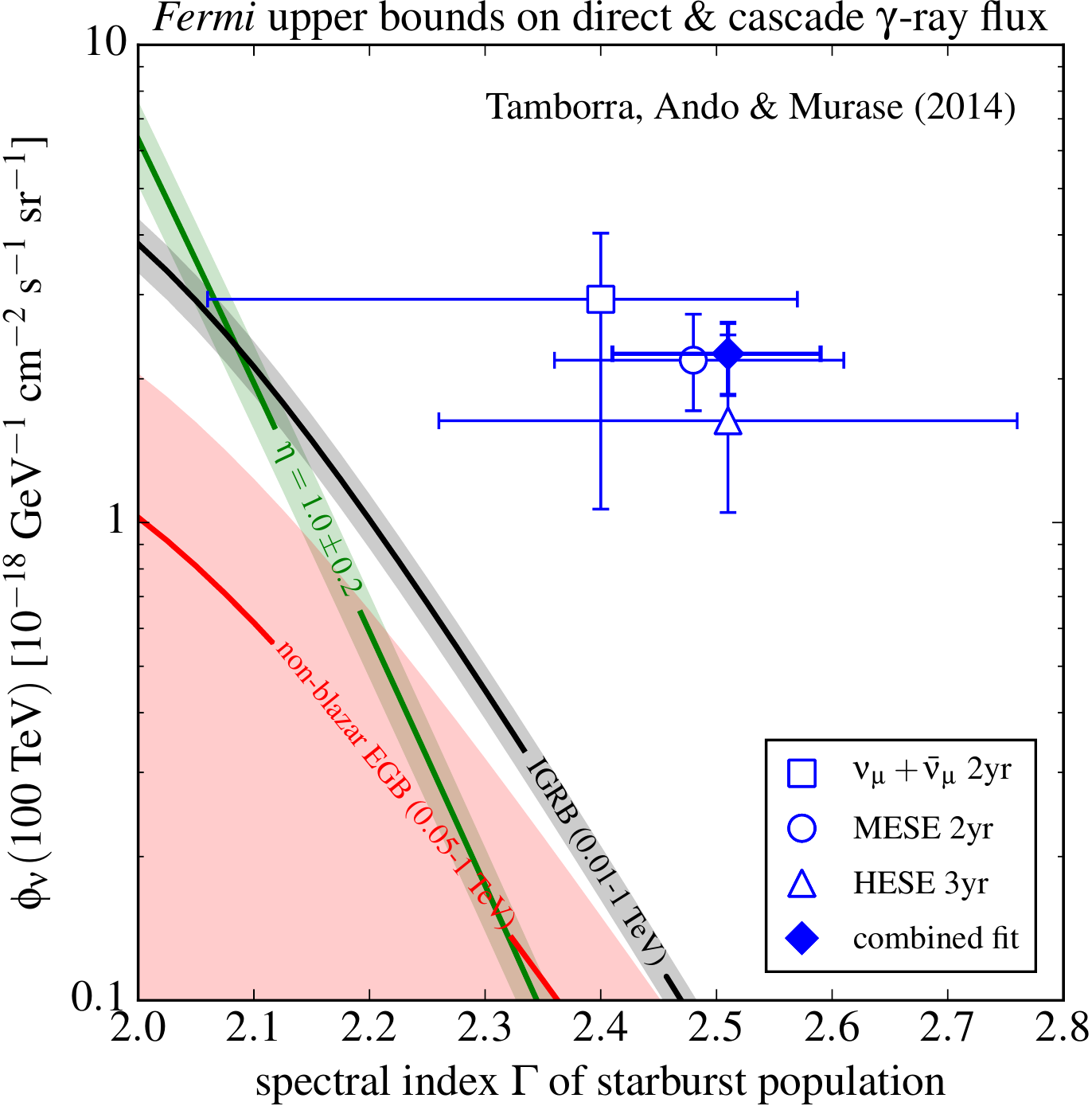}
\caption[]{Upper limits on the per-flavor normalization $\phi_\nu(100{\rm TeV})$ of \sfgs depending on the starburst spectral index $\Gamma_{\rm SB}$. The model of \cite{Tamborra:2014xia} is restricted to the green band where we allow for a $20\%$ uncertainty of the absolute normalization from the IR-\gammaRayHyph correlation. The black and red lines show the upper limits from the IGRB (0.01--1~TeV) and from the non-blazar EGB (0.05--1~TeV), respectively. Both results are shown with uncertainty bands. The data points show the best-fit power-law neutrino spectrum including the 68\% C.L.~range in terms of the spectral index $\Gamma$ and astrophysical normalization at 100~TeV estimated by IceCube analysis: the high-energy starting event (HESE) analysis~\citep{Aartsen:2014gkd}, the medium-energy starting event (MESE) analysis~\citep{Aartsen:2014muf} and the classical search for up-going $\nu_\mu+\bar\nu_\mu$ tracks~\citep{Aartsen:2015rwa}. The values are extracted from \cite{Aartsen:2015ita}, which also derives a combined fit to the data.}\label{fig3}
\end{figure}

\gammaRays and neutrinos are produced together in \sfgs via the production and decay of energetic pions from hadronic CR interactions. The two emission rates are related as
\begin{equation}\label{eq:Qgamma}
\frac{1}{3}\sum_{\alpha}E^2_\nu Q_{\nu_\alpha}(z,E_\nu) \simeq \frac{K_\pi}{4}E^2_\gamma Q_\gamma(z,E_\gamma)\,,
\end{equation}
where we introduce the relative charged-to-neutral pion rate $K_\pi$. For proton-gas ($pp$) collisions, we assume $K_\pi\simeq2$, corresponding to an equal contribution of $\pi^-$, $\pi^0$ and $\pi^+$. The average energies of \gammaRays and neutrinos are related as $E_\gamma\simeq 2E_\nu$. 

The corresponding diffuse flux of neutrinos observed at Earth is then given by the redshift integral
\begin{equation}
\phi_{\nu}(E_\nu) = \frac{c}{4\pi}\int_0^{z_{\rm max}}\frac{dz}{H(z)}Q_{\nu}(z,(1+z)E_\nu)\,,
\end{equation}
where $H(z)$ corresponds to the Hubble parameter at redshift $z$ and we assume maximum redshift of $z_{\rm max}=4$ in our calculations. 

In the case of \gammaRays, we must account for interactions with cosmic radiation backgrounds between the source and the observer. Pair production from \gammaRays via scattering off photons of the cosmic microwave background (CMB) peaks at PeV energies with an absorption length of only 10~kpc. Inverse-Compton scattering of high-energy electrons and positrons with the same photon background creates secondary high-energy \gammaRays that are again above the pair-production threshold. Therefore, the super-TeV electromagnetic energy gets quickly shifted into the sub-TeV range observable with {\it Fermi}. Whereas the CMB is the main driver of these electromagnetic cascades, the final spectrum depends also on pair-production on the extragalactic background light (EBL). In the following, we adopt the model of \cite{Dominguez:2010bv}, which provides tables of the EBL spectrum in the redshift range $0<z<4$.

The left panels of Figure~\ref{fig2} show the \gammaRayHyph and neutrino emission for the cases $\Gamma_{\rm SB}= \{2.0, 2.15, 2.3\}$ with $\eta=1$. 
The direct \gammaRayHyph and per-flavor neutrino predictions are in good agreement with the results shown in Figure~5 of \cite{Tamborra:2014xia}. 
Here, we also show the contribution from cascade \gammaRays that can enhance the overall emission if high-energy \gammaRays escape the galactic environment unattenuated.
This extra contribution, which was not included in the  \cite{Tamborra:2014xia} analysis, becomes important for hard emission ($\Gamma_{\rm SB}=2.0$) as shown in the top left panel of Figure~\ref{fig2}.
The right panels of Figure~\ref{fig2} show the required renormalization ($\eta<1$) that would saturate the best-fit non-blazar EGB constraint in the energy range 0.05--1~TeV.

In the hard-spectrum scenario with $\Gamma_{\rm SB}=2.0$, starbursts could explain the PeV neutrino data while satisfying the non-blazar EGB constraint if one allows for a rescaling of the hadronic emission by $\eta\simeq0.2$ compared to the IR-\gammaRayHyph luminosity correlation.
$\Gamma_{\rm SB}=2.0$ is also harder than the observed spectra of all \gammaRayHyph-detected starbursts, including the ultra-luminous infrared galaxy Arp~220 (Table~\ref{tab1}).
Even in this case, the neutrino data below 100~TeV exceed the prediction. 
On the other hand, a soft spectrum with $\Gamma_{\rm SB}=2.3$ is consistent with $\eta\simeq1$, the non-blazar EGB constraint, and the \gammaRayHyph spectra of individual starbursts, but the TeV--PeV neutrino flux is one order of magnitude below the IceCube signal. 

The predicted neutrino spectrum above $10$~TeV is dominated by the hard emission from starburst and SF-AGN galaxies and practically follows a power law with index $\Gamma_{\rm SB}$. We can therefore compare the high-energy tail of the neutrino emission to the best-fit power-law model of IceCube in the following. Figure~\ref{fig3} shows the scan of this model over different spectral indices $\Gamma_{\rm SB}$ and per-flavor neutrino flux normalizations at $100$~TeV. In this scan we allow the scaling factor $\eta$ to float in order to illustrate the tension with the neutrino observation. The black and red lines show the upper limits from the IGRB (0.01--1~TeV) and from the non-blazar EGB (0.05--1~TeV), respectively, within their uncertainty bands. The data points in Figure~\ref{fig3} show the best-fit power-law neutrino spectrum including the 68\% C.L.~range in terms of the spectral index $\Gamma$ and astrophysical normalization at 100~TeV estimated by IceCube analysis: the high-energy starting event (HESE) analysis~\citep{Aartsen:2014gkd}, the medium-energy starting event (MESE) analysis~\citep{Aartsen:2014muf} and the classical search for up-going $\nu_\mu+\bar\nu_\mu$ tracks~\citep{Aartsen:2015rwa}. The combined fit of this data is also shown as the filled data point~\citep{Aartsen:2015ita}.

The model of \cite{Tamborra:2014xia} with $\eta=1$ is indicated in Figure~\ref{fig3} as a green line, where we allow for a 20\% uncertainty on the normalization of the IR-\gammaRayHyph luminosity correlation \citep{Ackermann:2012vca}. As was already visible in Figure~\ref{fig2}, the non-blazar EGB constraint (within 68\% C.L.) requires softer emission with $\Gamma_{\rm SB} \gtrsim 2.15$. 
This index is also consistent with the \gammaRayHyph spectra of individual starburst galaxies summarized in Table~\ref{tab1}. In any case, the neutrino data in the 25~TeV--2.8~PeV energy range~\citep{Aartsen:2015ita} favors a softer power-law index and higher normalization than allowed by the non-blazar EGB constraint.

\section{Generic Cosmic Ray Calorimeters}
\label{sec:calorimeter}

In the previous section we examined the specific case of hadronic \gammaRayHyph and neutrino emission from \sfgs following the model of \cite{Tamborra:2014xia}. 
We now turn to the more general case of CR calorimeters, focusing on models motivated by the IceCube measurements.
Importantly, our generic CR calorimeter scenario is not based on multiwavelength scaling relations or luminosity functions, and could be applied to any population of hadronuclear ($pp$) neutrino sources that are optically thin to \gammaRays in the LAT energy range.

We approximate the cumulative neutrino spectrum (per flavor) of the population to follow a broken power law with an exponential cutoff:
\begin{equation}\label{eq:Nnu}
\frac{dN_\nu}{dE_\nu} \propto \begin{cases}E_\nu^{-2}&E_\nu<25{\rm TeV}\\E_\nu^{-\Gamma}&25{\rm TeV}<E_\nu<10{\rm PeV}\\E_\nu^{-\Gamma}e^{-E/{\rm 10 PeV}}&10{\rm PeV}<E_\nu.\end{cases}
\end{equation}
This model is designed to give a minimal contribution to the EGB at GeV energies, assuming that the parent CR spectral index below the break is $\Gamma \gtrsim 2$. 
The spectral break at 25~TeV is tuned to match the low-energy end of the neutrino data; a break at lower energies would increase the GeV \gammaRayHyph emission and the bounds from the non-blazar EGB would become stronger.
For this generic calorimeter model we assume that the emission rate can be expressed as the product $Q_{\nu}(z,E_\nu) \propto \rho(z)dN_{\nu}(E_\nu)/dE_\nu$ where $\rho(z)$ is the redshift evolution following the star formation rate in the redshift range $0<z<4$ \citep{Yuksel:2008cu}. 

\begin{figure*}[t]\centering
\includegraphics[width=0.45\linewidth]{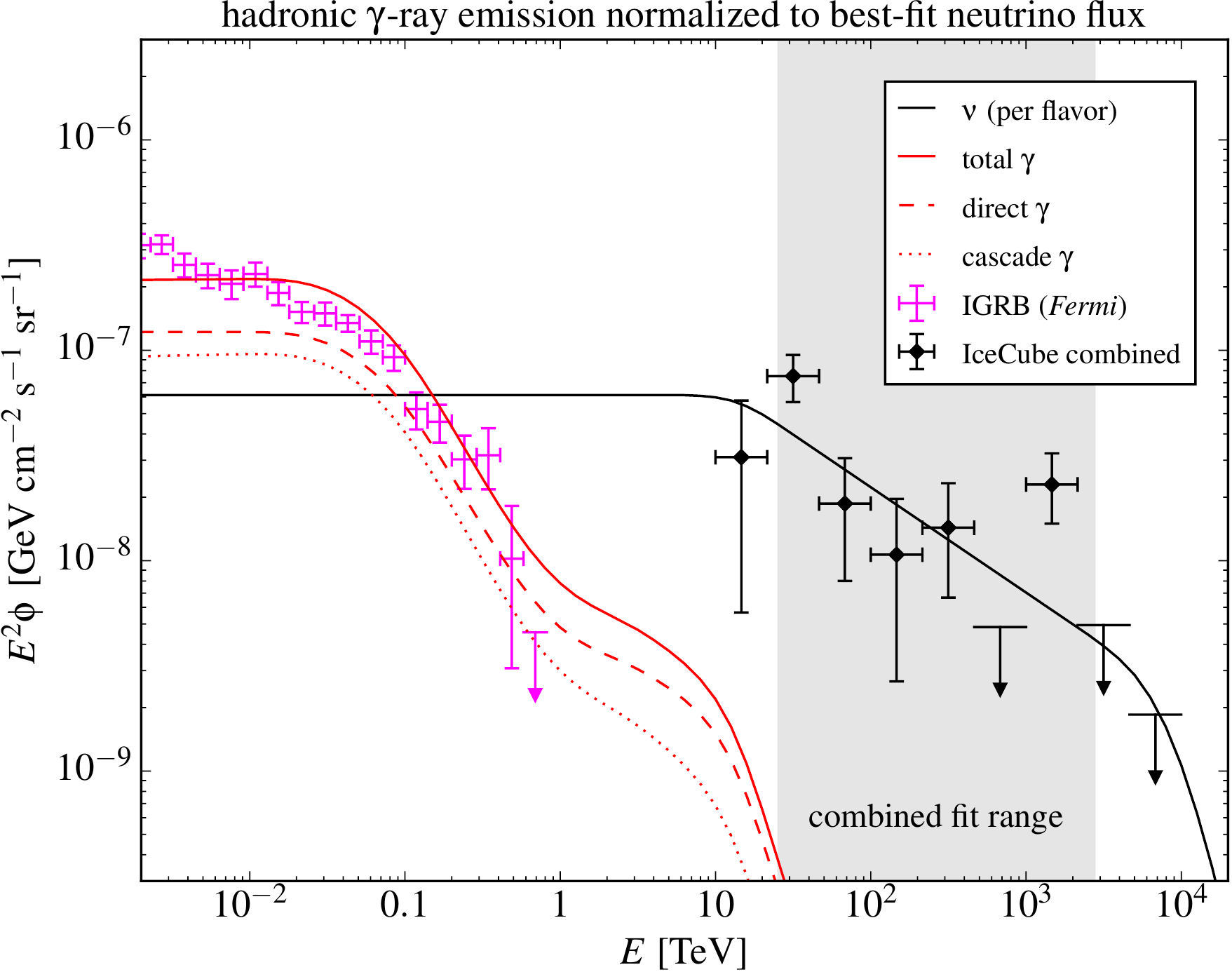}\hspace{0.3cm}
\includegraphics[width=0.45\linewidth]{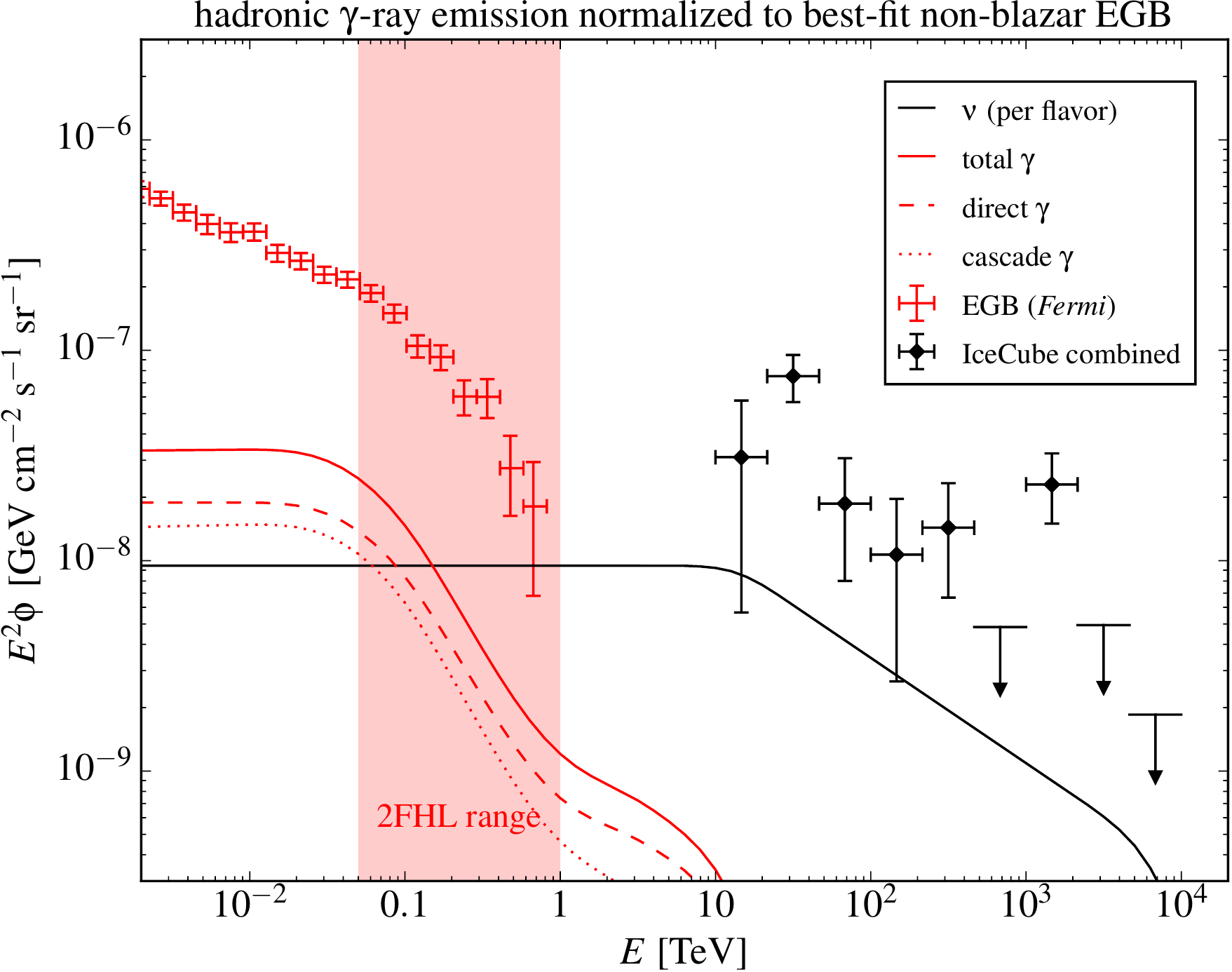}\\
\caption[]{The hadronic \gammaRayHyph (red lines) and per-flavor neutrino (black lines) contribution of generic CR calorimeters following the model of Equation~(\ref{eq:Nnu}) with $\Gamma=2.5$. We show the contribution of direct and cascaded \gammaRays separately as dashed and dotted lines, respectively. In the left plot the emission is normalized according to the best-fit of the combined neutrino data~\citep{Aartsen:2015ita} in the 25~TeV to 2.8~PeV energy range (gray-shaded area). The corresponding total \gammaRayHyph emission is only marginally consistent with the isotropic \gammaRayHyph background (IGRB). In the right plot we show the same model normalized to the best-fit 14\% non-blazar emission in the 0.05--1~TeV EGB (red-shaded area).}\label{fig4}
\end{figure*}

Figure~\ref{fig4} shows the contributions of diffuse neutrinos and the sum of direct and cascade \gammaRays for the emission spectrum of Equation~(\ref{eq:Nnu}). 
The left panel shows the normalization corresponding to the combined fit of neutrino data from 25~TeV to $2.8$~PeV (gray-shaded region) from \cite{Aartsen:2015ita}. 
We find that even for a fine-tuned spectrum with a break at 25~TeV, the hadronic \gammaRayHyph emission is only marginally consistent with the isotropic \gammaRayHyph background (IGRB). 
The right plot shows the same emission model, but normalized to the best-fit non-blazar contribution to the EGB in the 0.05--1~TeV energy range (red-shaded region). 
This new limit provides a stronger bound on the maximally allowed neutrino flux.

\begin{figure*}[t]\centering
\includegraphics[width=0.45\linewidth]{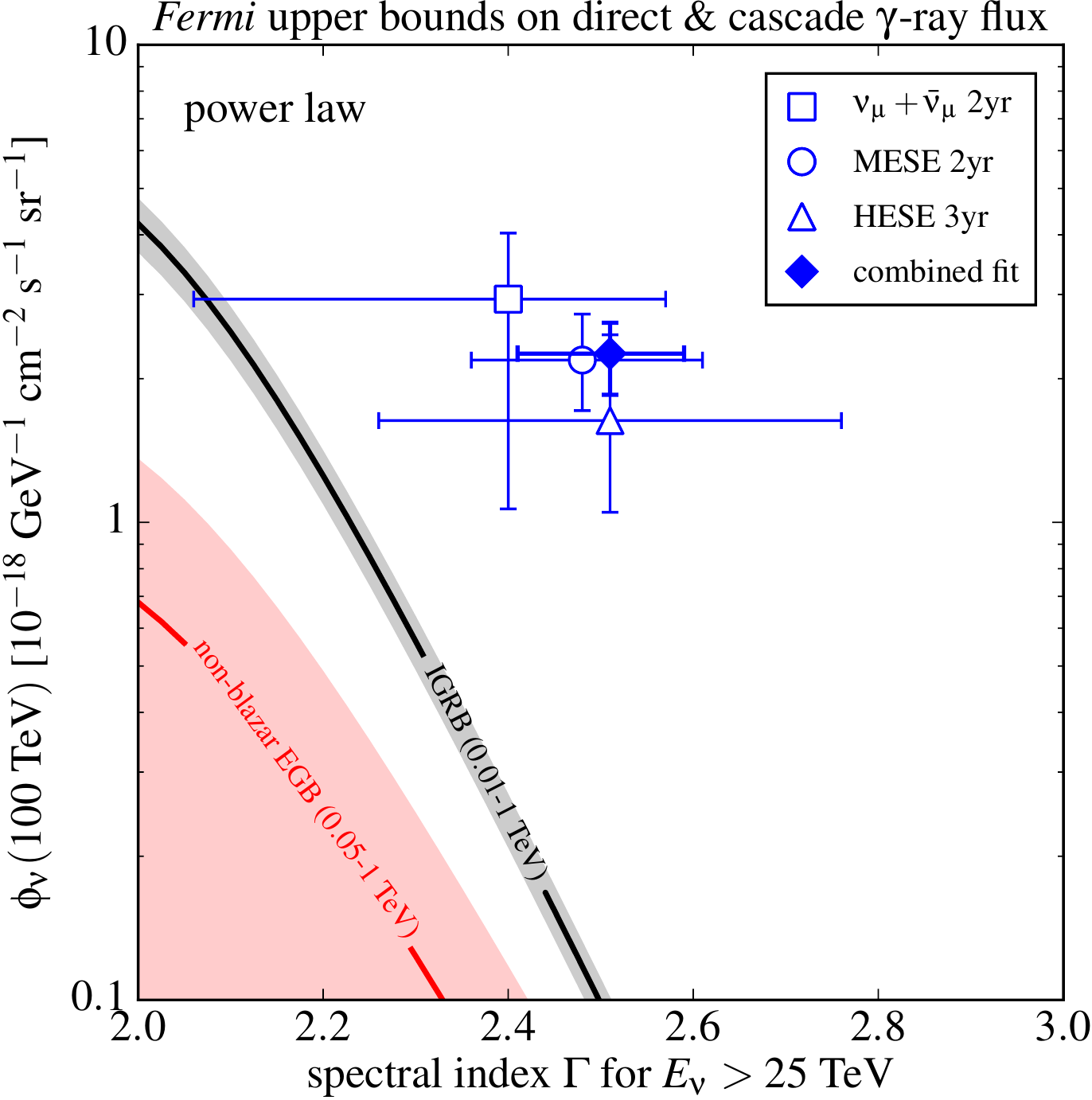}\hspace{0.3cm}\includegraphics[width=0.45\linewidth]{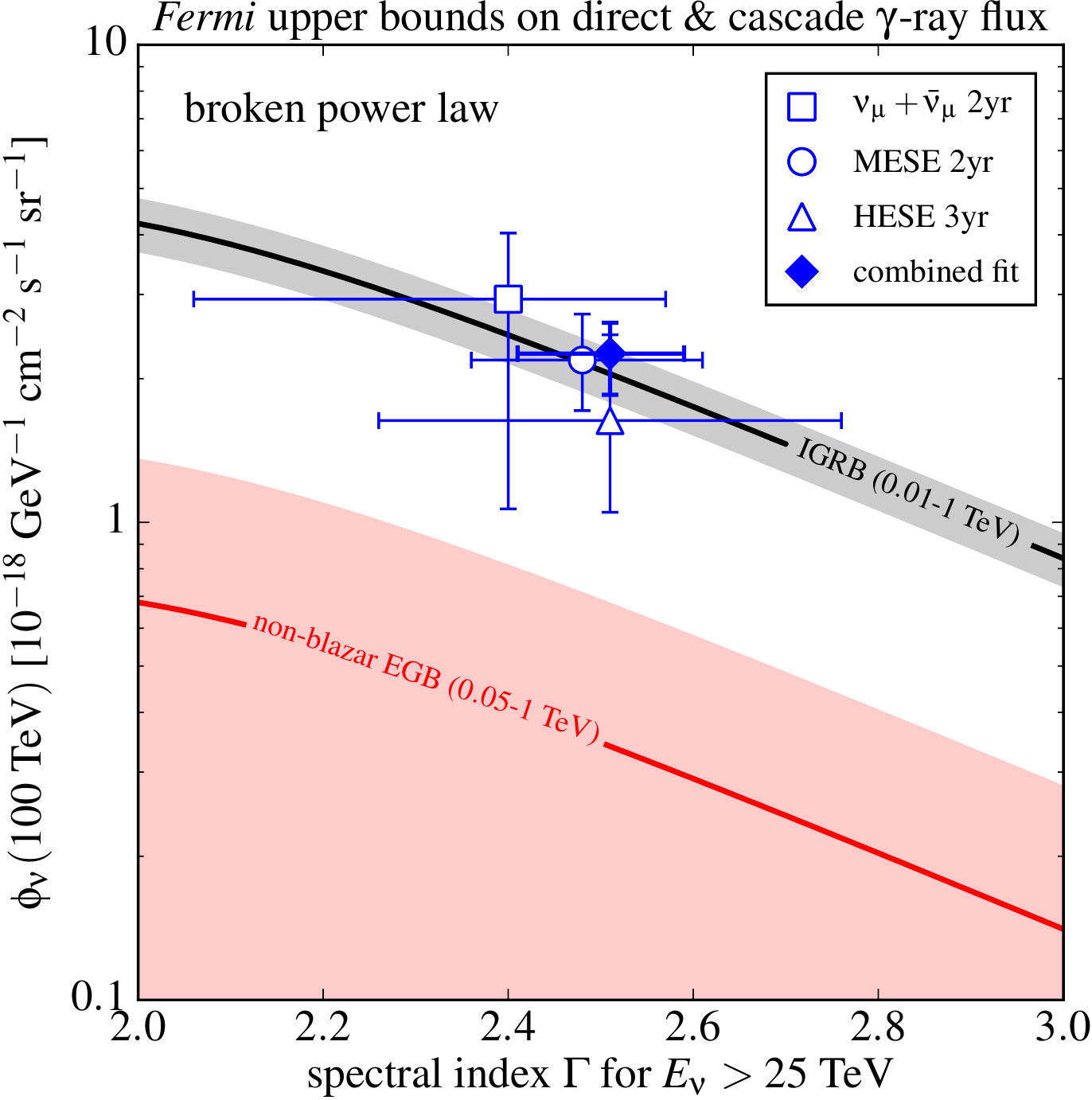}
\caption[]{Same as Figure~\ref{fig3}, but now showing the upper limits on the per-flavor normalization $\phi_\nu(100{\rm TeV})$ in terms of the spectral index $\Gamma$ of the high-energy neutrino spectrum for generic CR calorimeters. 
The left panel shows the constraints for a simple power-law emission spectrum and the right panel the constraints for a broken power-law model following Equation~(\ref{eq:Nnu}).}\label{fig5}
\end{figure*}

Figure~\ref{fig5} shows the limits on the neutrino flux normalization for different spectral indices $\Gamma$. 
The left panel shows results for a simple power-law model without a break at $25$~TeV. 
This scenario corresponds to the method of \cite{Murase:2013rfa} and \cite{Tamborra:2014xia} that derived strong limits on the spectral index ($\Gamma\lesssim2.2$) to explain the IceCube signal without overproducing the IGRB. 

For the broken power-law model of Equation~(\ref{eq:Nnu}) with a spectral break tuned to the low-energy end of the neutrino data at 25~TeV, the limits become weaker, as shown in the right panel of Figure~\ref{fig5}. 
Whereas the IGRB limit is marginally consistent with the neutrino data, the maximally allowed non-blazar EGB contribution places a strong constraint on this model. 

\section{General Considerations and Systematic Uncertainties}
\label{sec:systematic}

The constraints discussed in the previous two sections are conservative in several respects.
First, the broken power law model for generic CR calorimeters considered in Section \ref{sec:calorimeter} was specifically designed to account for the IceCube signal while producing a minimal contribution to the EGB.
As shown in Section \ref{sec:multi-messenger}, a realistic population of \sfgs would include non-negligible contributions from quiescent galaxies with softer spectral indices, in addition to the hard component from starbursts that is most relevant for the IceCube signal \citep{Tamborra:2014xia}. 
Second, any leptonic emission from \sfgs would result in additional \gammaRayHyph emission without a neutrino counterpart.
Finally, other extragalactic source populations, such as mis-aligned AGN are expected to have comparable EGB contributions to \sfgs \citep{Inoue:2011bm,DiMauro:2013xta,Hooper:2016gjy}.
Each of these factors would imply a more stringent upper bound on the cumulative hadronic \gammaRayHyph emission of \sfgs at energies above 50~GeV, and accordingly, a more stringent upper bound on their neutrino emission at TeV to PeV energies.

\begin{figure*}[p]\centering
\includegraphics[width=0.38\linewidth]{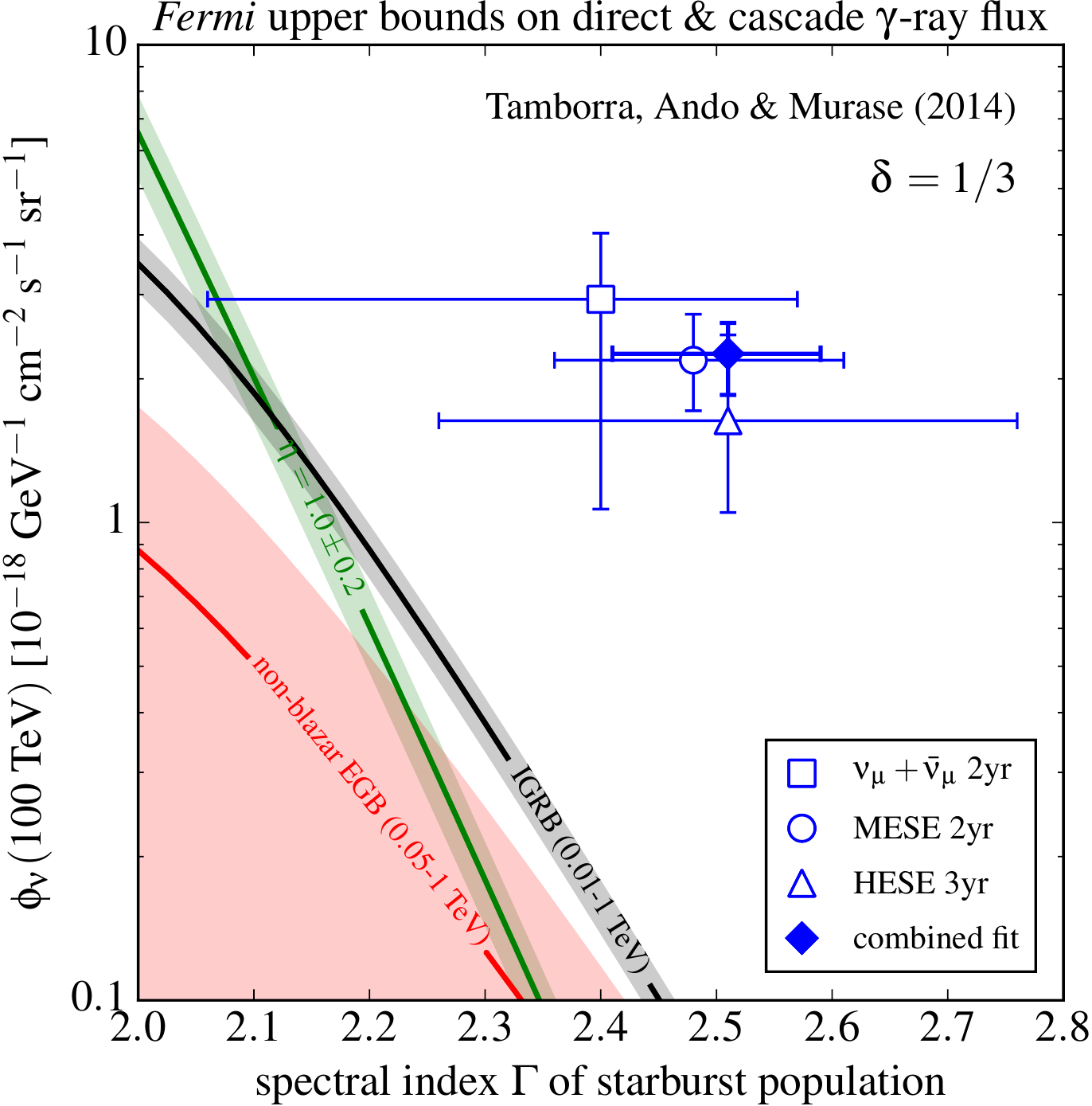}\hspace{0.3cm}\includegraphics[width=0.38\linewidth]{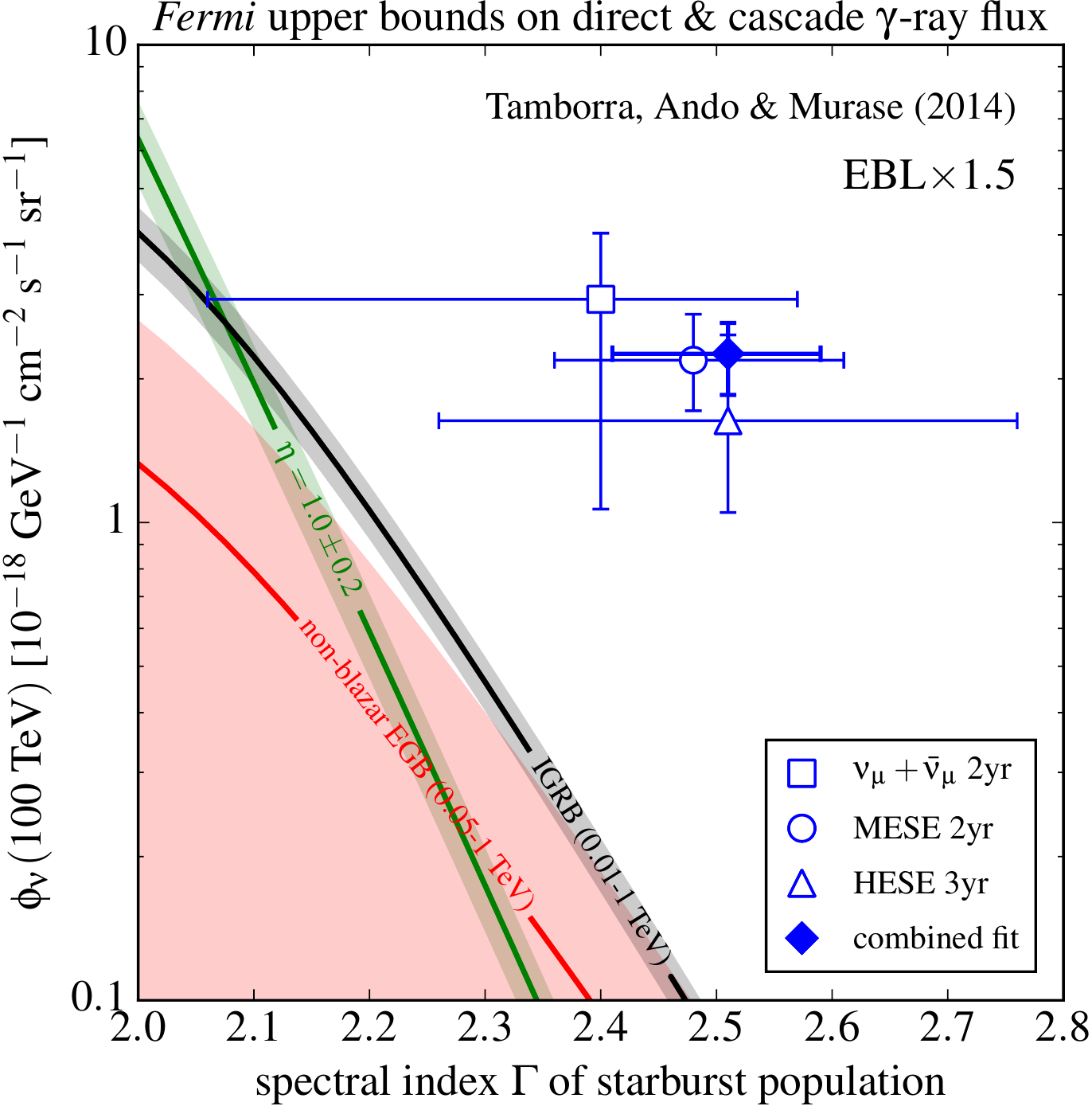}\\[0.3cm]
\includegraphics[width=0.38\linewidth]{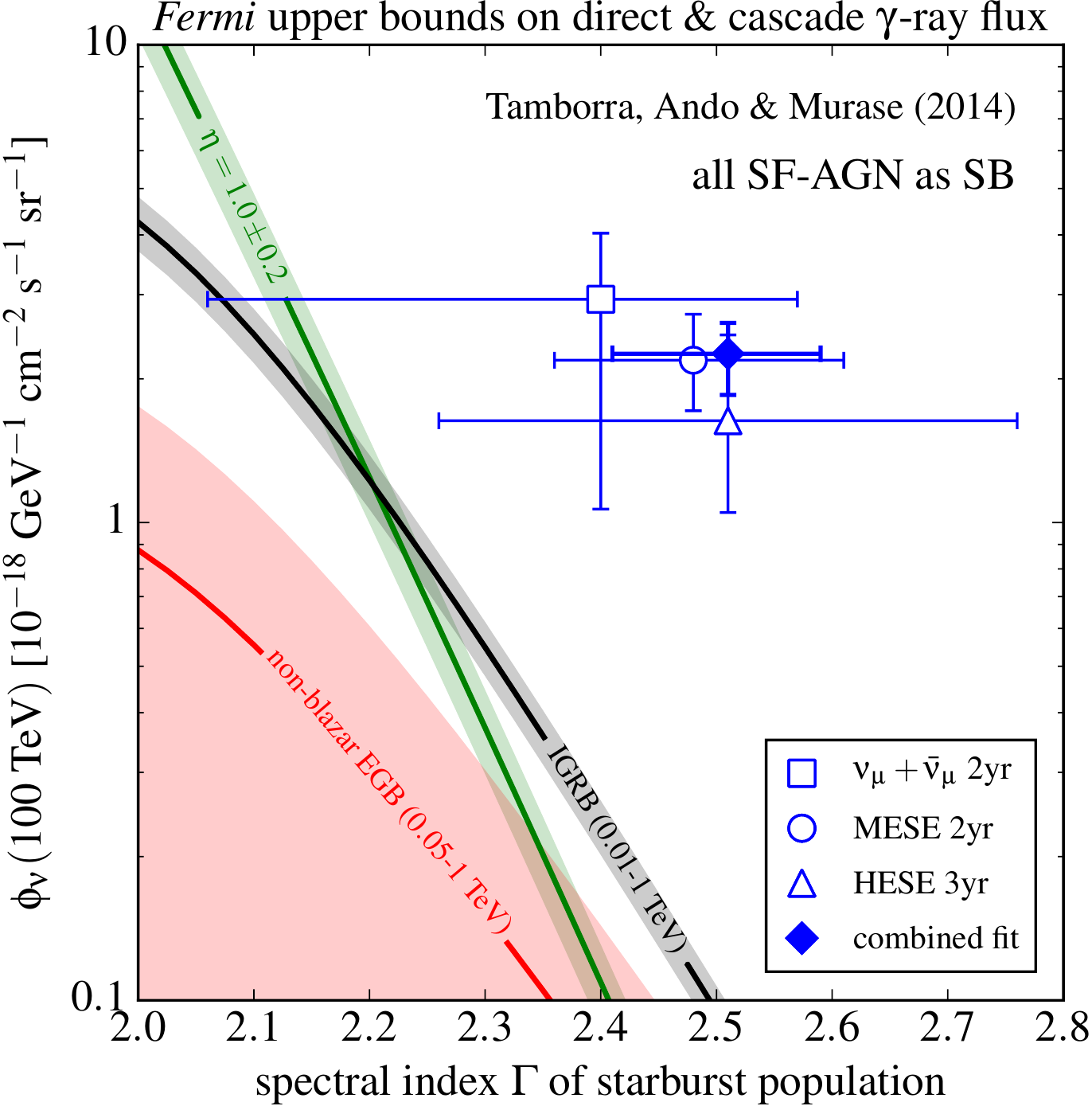}\hspace{0.3cm}\includegraphics[width=0.38\linewidth]{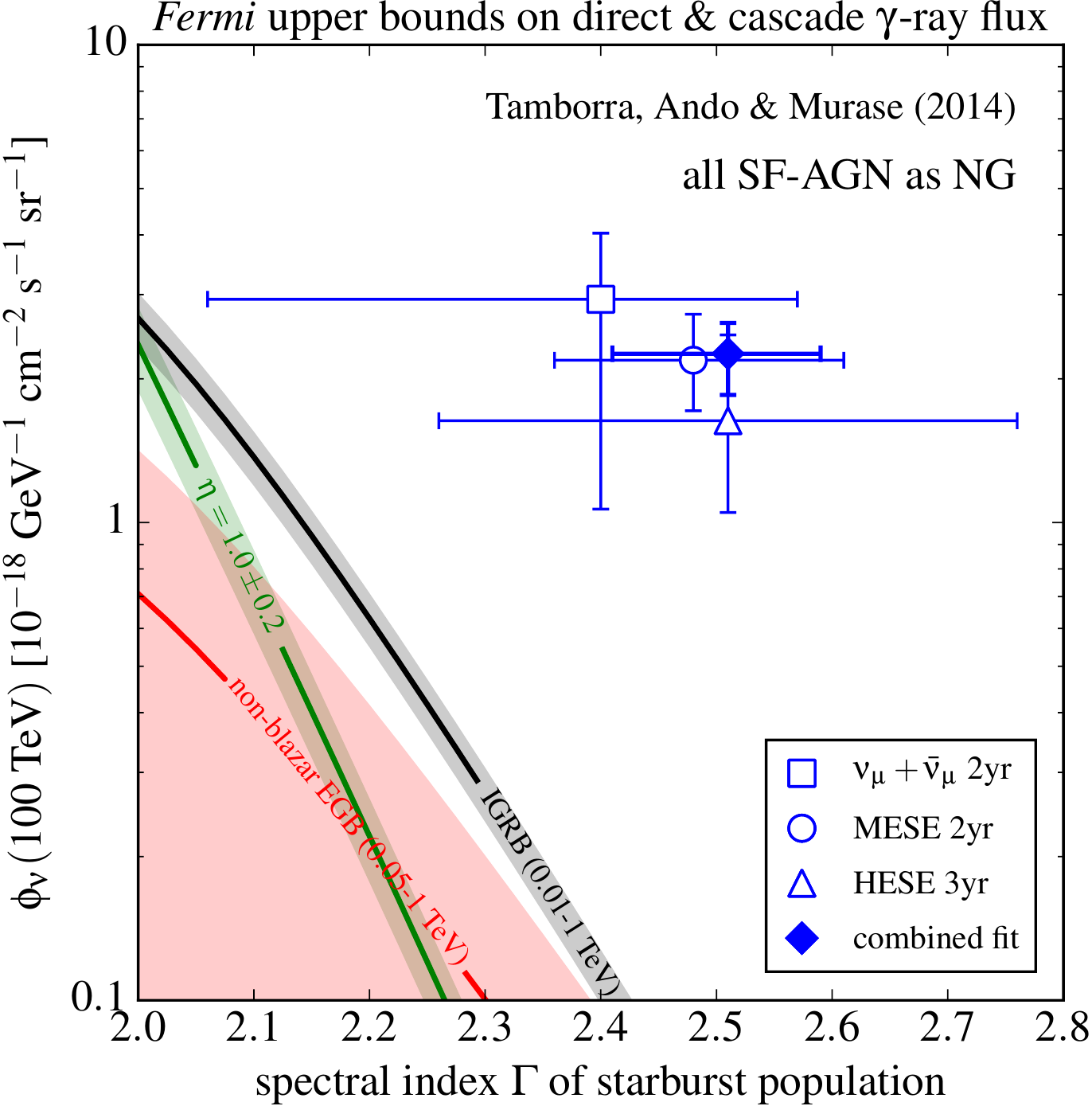}\\[0.3cm]
\includegraphics[width=0.38\linewidth]{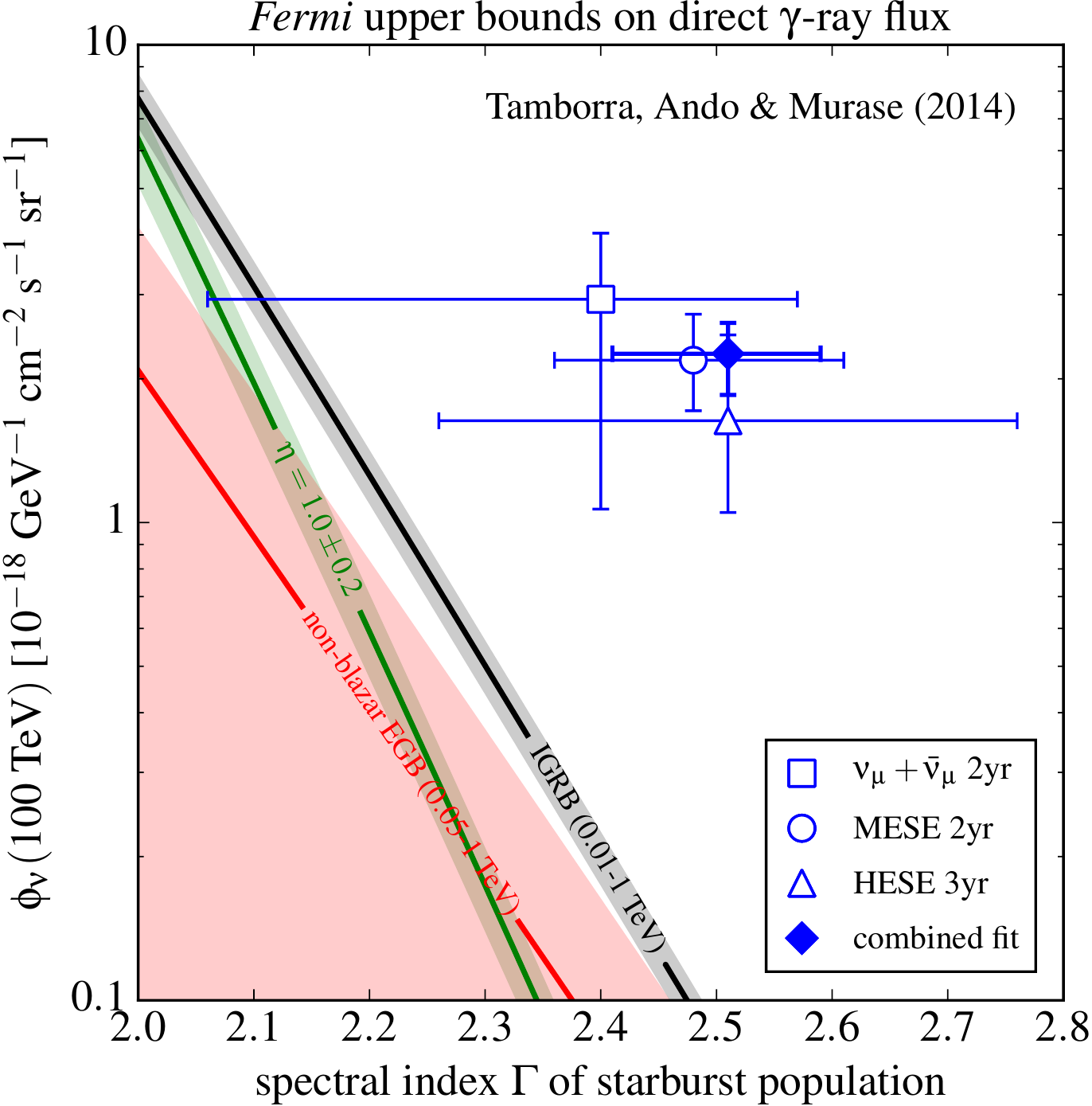}\hspace{0.3cm}\includegraphics[width=0.38\linewidth]{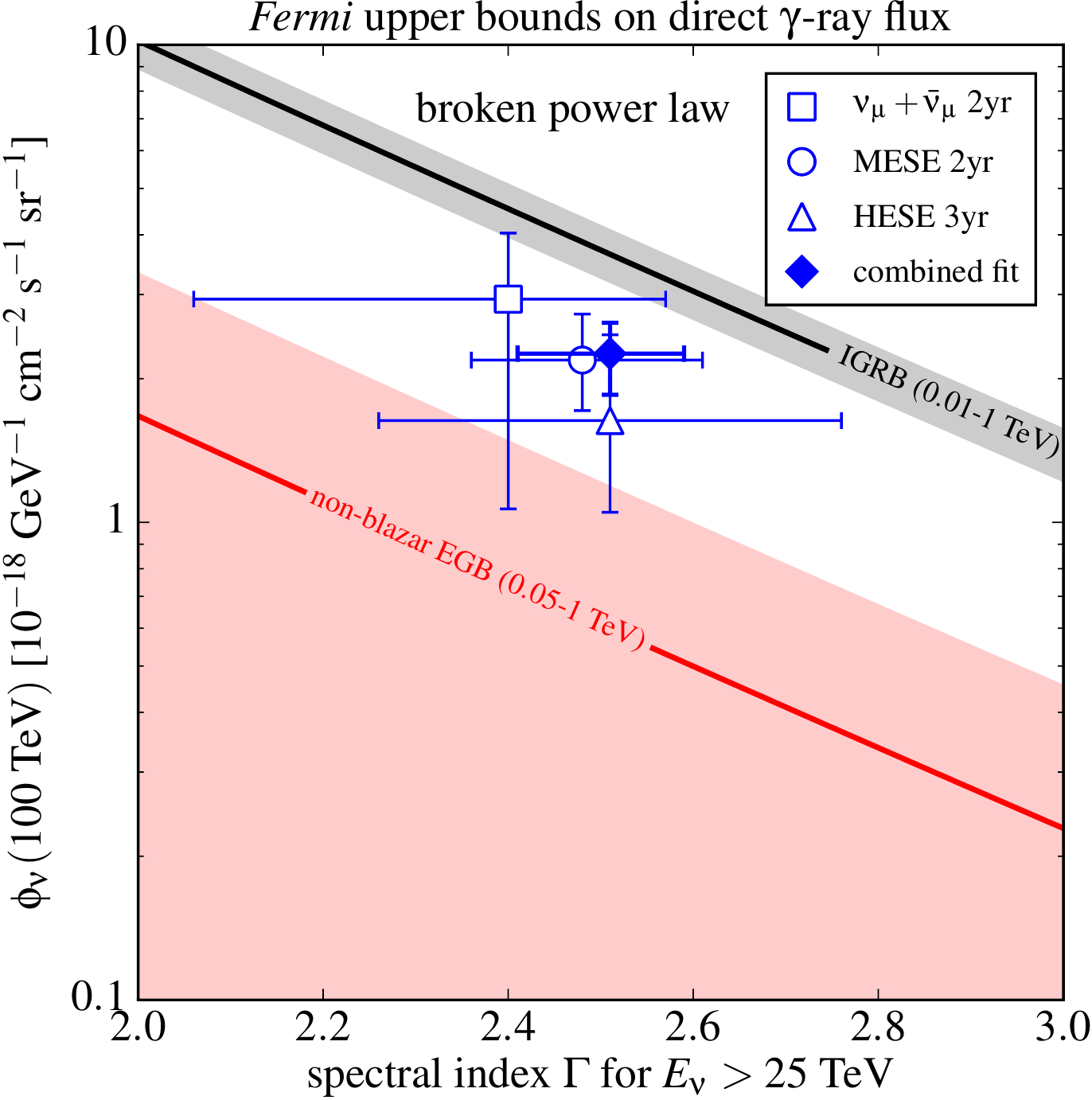}
\caption[]{The effect of model variations on the $\gamma$-ray bounds compared to the nominal results shown in Figs.~\ref{fig3} and \ref{fig5}. {\bf Top:} The \sfg model of \cite{Tamborra:2014xia} for a diffusion index $\delta=1/3$ (left) and for an extreme EBL intensity estimated by rescaling the model of \cite{Dominguez:2010bv} by a factor of $1.5$ (right). {\bf Middle:} The model of \cite{Tamborra:2014xia} assuming that all SF-AGN galaxies have hard emission as SB (left) or soft emission as NG (right). {\bf Bottom:} The model of \cite{Tamborra:2014xia} (left) and the generic model with spectrum (\ref{eq:Nnu}) (right) omitting the contribution of cascade \gammaRays.}\label{fig6}
\end{figure*}

We tested several variations to our fiducial models to explore the impacts of systematic uncertainties and changing different model parameters.
For the \cite{Tamborra:2014xia} model, we considered a variation with diffusion index $\delta=1/3$ instead of $\delta=1/2$ for normal galaxies, shown in the top left panel of Figure~\ref{fig6}.
The IGRB as well as non-blazar EGB bounds become slightly stronger in this case.

A higher EBL density at low redshift could deplete the $>50$~GeV \gammaRayHyph spectrum and reduce the tension with the non-blazar EGB bound.
We have checked that enlarging the EBL density by 50\% at all redshifts --- the maximum increase allowed by observations of individual blazars~\citep{Ackermann:2012sza} --- does not substantially affect our conclusions.
The corresponding bounds for the model of \cite{Tamborra:2014xia} are illustrated in the top right panel of Figure~\ref{fig6}.

A larger fraction of soft emitting SF-AGN galaxies could also reduce the tension with the non-blazar EGB constraint. The middle right panel of Figure~\ref{fig6} shows a calculation assuming that {\it all} SF-AGN galaxies in the model by \cite{Tamborra:2014xia} are treated as normal galaxies. Indeed, for this case the benchmark model $\eta=1$ is even compatible with the non-blazar EGB bound for a spectral index $\Gamma\simeq2.1$. However, with this model variation the predicted neutrino emission of \sfgs is still in strong tension with the neutrino data. On the other hand, the middle left panel of Figure~\ref{fig6} shows the case of treating all SF-AGN galaxies as SB galaxies with hard emission. This model variation increases the tension of the benchmark model with \gammaRayHyph bounds.

Even if we only consider contributions from direct \gammaRayHyph emission and neglect cascade contributions, the neutrino data is only marginally consistent with the non-blazar EGB limit, as shown in the bottom panels of Figure~\ref{fig6} for the \cite{Tamborra:2014xia} model (left), and the generic CR calorimeter model (right).
Suppression of the cascade component could result from $e^{+} e^{-}$ pair production in the intense radiation fields within the starburst itself that prevent very high energy photons from escaping \citep{Torres:2004ui,Inoue:2010iz,Chang:2014hua,Yoast-Hull:2015iea}.
However, the \gammaRayHyph opacity within starbursts is only expected to be significant at multi-TeV energies and is therefore not expected to strongly affect our constraints based on the cumulative emission of \sfgs above 50 GeV.
Alternatively, it has been suggested that plasma instabilities in intergalactic space might suppress cascade development \citep{Broderick:2011av}.

\section{Conclusions}
\label{sec:discussion}

New studies of the EGB composition at energies above 50~GeV find a dominant contribution from blazars, leaving only a $\sim14$\% residual component attributed to all other source classes, including \sfgs. 
Motivated by this bound, we studied the cumulative hadronic \gammaRayHyph and neutrino emission of \sfgs.
Figure~\ref{fig3} summarizes our main result that \sfgs are now disfavored as a dominant component of the IceCube astrophysical neutrino signal. 
Hadronic emission from \sfgs that is consistent with both the IR-\gammaRayHyph luminosity correlation and the non-blazar EGB bound requires soft emission with $\Gamma_{\rm SB}\gtrsim2.15$, matching the observed spectra of individual \gammaRayHyph-detected starburst galaxies. 
Taking the $\Gamma_{\rm SB}=2.2$ case as an example, the maximal contribution of \sfgs to the best-fit diffuse neutrino background of~\citet{Aartsen:2015ita} is $\sim30\%$ at 100~TeV and $\sim60\%$ at 1 PeV when saturating the upper bound (28\%) on the non-blazar fraction of the EGB.

We have also studied the emission of generic CR calorimeters, allowing for hard \gammaRayHyph emission below 25~TeV to avoid the non-blazar EGB limit. 
These results are summarized in Figure~\ref{fig5}. 
Following the model of Equation~(\ref{eq:Nnu}) and assuming the best-fit normalization and spectral index of~\citet{Aartsen:2015ita}, the maximal contribution of \sfgs to the diffuse neutrino background between 25~TeV and 2.8~PeV is $\sim30\%$, again saturating the upper bound on the non-blazar EGB component.

The astrophysical neutrino signal reported by IceCube is the component that remains after accounting for atmospheric backgrounds, which are increasingly important towards lower energies.
If the IceCube signal were substantially contaminated by unaccounted atmospheric backgrounds, our constraints on the relative contribution of \sfgs to astrophysical neutrino emission would be weakened, while the absolute limits on their neutrino emission would be unchanged.
However, multiple empirical \cite[e.g.,][]{Aartsen:2014muf} and theoretical \cite[e.g.,][]{Halzen:2016thi} arguments disfavor this scenario, and a deep study of atmospheric backgrounds in IceCube is beyond the scope of this work.

We draw two main conclusions from the results above:

{\it (i)} The high-energy neutrino emission of several of the most prominent non-thermal extragalactic source classes is now bounded by an ensemble of multi-messenger constraints. A joint-likelihood search targeting \gammaRayHyph blazars finds that this population can account for $<$19--27\% of the IceCube flux \citep{Aartsen:2016lir}. A similar search towards GRBs excludes a contribution larger than 1\% \citep{Aartsen:2014aqy}. In this work, we argue that a third class of extragalactic sources, \sfgs, is also likely a sub-dominant component. Together, these bounds imply that the sources of high-energy IceCube neutrinos are not readily traced by extragalactic \gammaRayHyph emitters, with the possible exception of radio galaxies. Given the tight expected connection between neutrino and \gammaRayHyph emission, one possibility is that the neutrinos originate from environments with high \gammaRayHyph opacity \citep{Murase:2015xka}, or that the neutrinos mainly come from entirely different source classes. For example, the above constraints would be alleviated if a large Galactic contribution were present, although many of these scenarios are also disfavored \citep{Ahlers:2015moa}.  
It is also possible that multiple distinct source classes have leading contributions over different parts of the TeV to PeV energy range.

{\it (ii)} An upper bound on the emission of \sfgs may be encouraging for those seeking the first individual high-energy neutrino sources. Starburst galaxies are among the most numerous candidate neutrino sources (local density of $\sim10^{-4}$ Mpc$^{-3}$) and therefore must be individually faint in order not to overproduce the measured neutrino flux. Given this high local density, and accounting for cosmic evolution, the cumulative emission of \sfgs is predicted to be nearly isotropic even on small angular scales \citep{Ahlers:2014ioa}. Also, the neutrino emission of individual \sfgs is expected to be steady over Myr timescales given the lifetime of CRs in the interstellar medium, and therefore no distinctive signatures in the time domain are available to enhance sensitivity to individual sources. For these reasons, if \sfgs were the main component of the diffuse neutrino background, the prospects for detecting individual neutrino sources would be rather bleak, requiring an exposure substantially larger than can be achieved with IceCube or even proposed next-generation neutrino telescopes such as IceCube-Gen2~\citep{Aartsen:2014njl} and KM3NeT~\citep{Adrian-Martinez:2016fdl}. The present results largely exclude that scenario, and therefore keep open the possibility that source classes with more conspicuous small-angle anisotropy signals and/or temporal variations may be found in the near future.

{\it Note Added.---}After the submission of this work, \citet{Kistler:2015ywn} also pointed out the difficulty of reconciling the high neutrino intensity observed at TeV energies and the limits set by the EGB for extragalactic sources.

{\it Acknowledgments.---}We acknowledge helpful discussions with Kohta Murase and Markus Ackermann.
We also thank two anonymous referees who encouraged us to broaden the scope of this work to include a more realistic \sfg population model.
A third referee gave constructive feedback on the presentation of these results.
MA is supported by the National Science Foundation under grants OPP-0236449 and PHY-0236449.

\bibliographystyle{apj}
\bibliography{main}

\end{document}